\documentclass[useAMS,usenatbib]{mn2e}
\usepackage{natbib}
\usepackage{amsmath}
\usepackage{subfig}
\usepackage{xcolor}

\newcommand*{\Msun}{\ensuremath{\mathrm{M_\odot}}}%

\usepackage{manfnt}
\usepackage{marvosym}
\usepackage{graphicx}
\usepackage{epstopdf}
\title[GAMA: Star Formation Histories]{Galaxy And Mass Assembly (GAMA): Linking Star Formation Histories and Stellar Mass Growth}

\author[A. E. Bauer et al.]{Amanda E. Bauer,$^{1,\triangle,}$\thanks{email: amanda.bauer@aao.gov.au}
Andrew~M.~Hopkins,$^{1}$
Madusha~Gunawardhana,$^{1}$
\newauthor Edward~N.~Taylor$^{2,3}$
Ivan~Baldry,$^{4}$ 
Steven~P.~Bamford,$^{5}$ 
Joss~Bland-Hawthorn,$^{2}$
\newauthor Sarah~Brough,$^{1}$  
Michael~J.~I.~Brown,$^{6}$
Michelle~E.~Cluver,$^{1,\triangle}$
Matthew~Colless,$^{1,7}$
\newauthor Christopher~J.~Conselice,$^{5}$ 
Scott~Croom,$^{2}$
Simon~Driver,$^{8,9}$
Caroline~Foster,$^{10}$
\newauthor D.~Heath~Jones,$^{6}$ 
Maritza~A.~Lara-Lopez,$^{1,\triangle}$
Jochen~Liske,$^{11}$
\newauthor  \'Angel R. L\'opez-S\'anchez,$^{1,12}$ 
Jon~Loveday, $^{13}$
Peder~Norberg,$^{14}$ 
Matt~S.~Owers,$^{1,\triangle}$
\newauthor Kevin~Pimbblet,$^{6}$
Aaron~Robotham,$^{8,9}$
Anne~E.~Sansom,$^{15}$
Rob~Sharp$^{7}$ \\
$^{1}$Australian Astronomical Observatory, PO Box 915, North Ryde, NSW 1670, Australia\\
$^{ \triangle}$Australian Research Council (ARC) Super Science Fellow\\
$^{2}$Sydney Institute for Astronomy (SIfA), School of Physics, University of Sydney, NSW 2006, Australia \\
$^{3}$School of Physics, The University of Melbourne, Parkville, VIC 3010, Australia\\
$^{4}$Astrophysics Research Institute, Liverpool John Moores University, Twelve Quays House, Egerton Wharf, Birkenhead, CH41 1LD, UK \\
$^{5}$School of Physics \& Astronomy, The University of Nottingham, University Park, Nottingham, NG7 2RD, UK \\ 
$^{6}$School of Physics, Monash University, Clayton, Vic 3800, Australia\\
$^{7}$Research School of Astronomy \& Astrophysics, The Australian National University, Cotter Road, Weston Creek, ACT 2611, Australia \\
$^{8}$School of Physics \& Astronomy, University of St Andrews, North Haugh, St Andrews, KY16 9SS, UK\\
$^{9}$ICRAR, The University of Western Australia, 35 Stirling Highway, Crawley, WA 6009, Australia\\
$^{10}$European Southern Observatory, Alonso de Cordova 3107, Vitacura, Santiago, Chile \\
$^{11}$European Southern Observatory, Karl-Schwarzschild-Str. 2, 85748 Garching, Germany\\
$^{12}$Department of Physics and Astronomy, Macquarie University, NSW 2109, Australia \\
$^{13}$Astronomy Centre, University of Sussex, Falmer, Brighton BN1 9QH \\
$^{14}$Institute for Computational Cosmology, Department of Physics, Durham University, Durham DH1 3LE, UK \\
$^{15}$University of Central Lancashire, Preston, Lancs, PR1 2HE, UK \\
}

\begin{document}

\date{Accepted -- . Received --}

\maketitle

\begin{abstract}
We present evidence for stochastic star formation histories in low-mass (M$_{*} < 10^{10}$ $\Msun$) galaxies from observations within the Galaxy And Mass Assembly (GAMA) survey.  For $\sim$73,000 galaxies between $0.05<z<0.32$, we calculate star formation rates (SFR) and specific star formation rates (SSFR = SFR/M$_{*}$) from spectroscopic H$\alpha$ measurements and apply dust corrections derived from Balmer decrements.  We find a dependence of SSFR on stellar mass, such that SSFRs decrease with increasing stellar mass for star-forming galaxies, and for the full sample, SSFRs decrease as a stronger function of stellar mass.  We use simple parametrizations of exponentially declining star formation histories to investigate the dependence on stellar mass of the star formation timescale and the formation redshift.  We find that parametrizations previously fit to samples of $z\sim1$ galaxies cannot recover the distributions of SSFRs and stellar masses observed in the GAMA sample between $0.05<z<0.32$.   In particular, a large number of low-mass (M$_{*} < 10^{10}$ $\Msun$) galaxies are observed to have much higher SSFRs than can be explained by these simple models over the redshift range of $0.05<z<0.32$, even when invoking mass-dependent staged evolution.  For such a large number of galaxies to maintain low stellar masses, yet harbour such high SSFRs, requires the late onset of a weak underlying exponentially declining SFH with stochastic bursts of star formation superimposed.
\end{abstract}


%%%%%%%%%%%%%%%%%%%%%%%%%%%%%%%%%%%%%%%%%%%%%%
%%                    INTRO                 %%
%%%%%%%%%%%%%%%%%%%%%%%%%%%%%%%%%%%%%%%%%%%%%%
\section{Introduction}\label{sec:intro}
 
 % ascii abstract
 %We present evidence for stochastic star formation histories in low-mass (M* < 10^10 Msun) galaxies from observations within the Galaxy And Mass Assembly (GAMA) survey.  For ~73,000 galaxies between 0.05<z<0.32, we calculate star formation rates (SFR) and specific star formation rates (SSFR = SFR/M*) from spectroscopic Halpha measurements and apply dust corrections derived from Balmer decrements.  We find a dependence of SSFR on stellar mass, such that SSFRs decrease with increasing stellar mass for star-forming galaxies, and for the full sample, SSFRs decrease as a stronger function of stellar mass.  We use simple parametrizations of exponentially declining star formation histories to investigate the dependence on stellar mass of the star formation timescale and the formation redshift.  We find that parametrizations previously fit to samples of z~1 galaxies cannot recover the distributions of SSFRs and stellar masses observed in the GAMA sample between 0.05<z<0.32.   In particular, a large number of low-mass (M* < 10^10 Msun) galaxies are observed to have much higher SSFRs than can be explained by these simple models over the redshift range of 0.05<z<0.32, even when invoking mass-dependent staged evolution.  For such a large number of galaxies to maintain low stellar masses, yet harbour such high SSFRs, requires the late onset of a weak underlying exponentially declining SFH with stochastic bursts of star formation superimposed.

As surveys of galaxy populations at high redshifts progress, it becomes increasingly important to understand how observed properties of galaxies at high redshift map onto those in the low redshift universe.  Simulations of structure formation over the history of the Universe have been impressively successful at predicting the growth of dark matter haloes and distributions of galaxies in space \citep{Springel2006,Croton2006}, but the challenge remains to determine the associated mass assembly and star formation within galaxies over time.   

Two main methods have been used to measure the cumulative growth of stellar mass as a function of epoch.  One approach is to look at the assembled stellar mass as a function of time \citep{Cole2001,Bell2003,Dickinson2003,Fontana2003,Drory2005,Bundy2006,PG2008,Wilkins2008,Li2009,Ilbert2010}, as shown by galaxy stellar mass functions \citep[e.g.][]{Baldry2008,Drory2009,Bolzonella2010,Marchesini2012}.   Recent surveys have shown that massive galaxies are already in place at early times and very large populations of low mass galaxies exist at all redshifts.  Deep surveys reaching masses below $10^{10} \Msun$ have also shown galaxy stellar mass functions with steep slopes at the low-mass end \citep[e.g.][]{Baldry2008,Drory2009,Pozetti2010,Baldry2012}, emphasising a need to understand the growth of these galaxies, the most populous in the universe.

The other main method for understanding galaxy assembly measures the instantaneous level of star formation to determine how much gas is being converted to stellar mass at any given time.  Observations spanning the electromagnetic spectrum reveal increasing SFRs with redshift at any given stellar mass \citep{Lilly1996,Cowie1996,Bauer05,Noeske2007_MS,Elbaz2007,Daddi2007,Caputi2008,Drory2008,Santini2009,Rujopakar2010,Oliver2010,Wuyts2011,Karim2011,Fontanot2012}, with evidence of a peak in global star formation activity between $1.5<z<3$ \citep[e.g.][]{Madau1996,HB2006,Bauer2011_GNS}.  

Studies of the star formation rates of galaxies have found varying results depending on factors like the redshift coverage, stellar mass range, method for determining SFR and the presence of dust.  Most studies use star-forming galaxies and find increasing SFRs with stellar mass \citep[e.g.][]{Noeske2007_MS,Santini2009,Bauer2011_GNS}, but this relationship is not as clear when samples start to include lower mass galaxies \citep{Leitner2012}, stacking measurements for SFRs \citep{Karim2011}, or to detect very low levels of star formation \citep{Wilman2008}.   Also, a positive correlation has been found between SFR and dust extinction by \citet{Zahid2013} using the SDSS DR7, which is consistent with the findings of \citet{Whitaker2012} who study the relationship to $z=2.5$. 

A standing issue with understanding galaxy growth is that  current models predict a different rate of growth in the stellar mass in galaxies than what is determined from the observed properties of galaxies \citep{Bower2006,DLB2007,Dave2008,Damen2009,Gilbank2011}.  Models have difficulty reproducing the distribution of SFRs seen in galaxies at a given stellar and cosmic time \citep{Fontanot2012}, which could involve various feedback prescriptions and other physical factors that regulate the timescales over which galaxies form stars.  \citet{Wuyts2011} find that star formation histories (SFH) for galaxies out to $z=3$ vary on long timescales compared to their dynamical times.  On the other hand, \citet{Caputi2008} are unable to fit $z<1$ galaxies with constant SFH and favour secondary bursts for massive galaxies. \citet{Drory2008} show that up to $z=5$, more massive galaxies have steeper and earlier onsets of star formation, and higher peak SFRs followed by a shorter decay time.  Most studies are of high mass galaxies, with $0.5<z<3$, using SDSS, which has a median redshift of $z=0.1$ as a local comparison.   Or at the other extreme, studies look at colour-magnitude diagrams of resolved stellar populations in dwarf galaxies in the local volume, which reveal that these galaxies have very complex SFHs \citep{Weisz2011}.  

To address this issue, we investigate how star formation varies as a function of stellar mass to gain insight as to how much of the stellar mass in galaxies is accumulated from in-situ star formation, rather than accreted from mergers.  With this paper, our goal is to use the unique spectroscopic coverage of the large Galaxy And Mass Assembly (GAMA) survey \citep{Baldry2010,Robotham2010,Driver2011} to examine the SSFR behaviour of a large sample of galaxies over a broad stellar mass and redshift range.  GAMA is complete to $r_{\rmn{AB}}=19.4$ mag for the redshift range $0.05<z<0.32$ over 144 square degrees.  We look at H$\alpha$-derived and dust-corrected SFRs to see how the SSFR up to $z=0.32$ evolves as a function of stellar mass and redshift.  

We also investigate how the stellar masses and SSFRs for GAMA galaxies compare to the prediction from simple exponentially declining star formation history models advocated by studies based on observations at $z\sim 1$ \citep{Noeske07_tau,Martin07,Gilbank2011,Bauer2011_GNS}.  We draw conclusions from the existence of the populations observed within the GAMA survey.  We are not complete to a given stellar mass for low- or non-star-forming galaxies, so we explore the SFH and mass assembly properties of this unique redshift sample.   The conclusions prove to be quite interesting using minimal constraints on our selection criteria, despite the unavoidable constraints of all magnitude-limited spectroscopic samples.  

Throughout the paper we assume the standard $\Lambda$CDM cosmology, a flat universe with $\Omega_\Lambda = 0.70$,  $\Omega_M = 0.30$ and a Hubble constant of $H_0 = 70 $ km s$^{-1}$ Mpc$^{-1}$.

%%%%%%%%%%%%%%%%%%%%%%%%%%%%%%%%%%%%%%%%
%%              GAMA DATA                  %%
%%%%%%%%%%%%%%%%%%%%%%%%%%%%%%%%%%%%%%%%

\section{Data}\label{sec:data}

The primary sample we use in this study is spectroscopic data from phase 1 of the Galaxy And Mass Assembly (GAMA-I) survey to redshift $z=0.32$.  Details of the GAMA survey can be found in \citet{Driver2011} with the survey input catalogue described in \citet{Baldry2010}, and the spectroscopic analysis and measurements detailed in {\citet{hopkins2013}.

The GAMA data presented in this work include three equatorial regions of the sky centred	at 09h, 12h and 14h30m, each covering 12$\times$4 degrees, giving a total area of 144 square degrees.  In this study, we are more than 98\% spectroscopically complete to the survey depth of $r_{AB} = 19.4$ \citep{Driver2011}.  We only include galaxies with a redshift quality indicator ${\rm nQ}\ge 3$ \citep{Driver2011}.  

In addition to the GAMA galaxies identified above, our main sample also includes all Sloan Digital Sky Survey (SDSS) galaxies observed within the GAMA volume.  These SDSS galaxies were not targeted for GAMA spectroscopy since they are brighter than $r_{AB} = 17.77$ and already have SDSS spectra available.  Emission line measurements, redshifts and stellar masses for these SDSS objects are obtained from the 7th SDSS Data Release (DR7) \citep{DR72009}.   All masses, star formation rates (SFRs) and dust corrections are calculated identically to the GAMA galaxies, as described below in Section~\ref{sec:sfr}.   

The total number of galaxies in the main sample used in this study, to the survey depth of $r_{AB} = 19.4$ and within the redshift range of $0.05<z<0.32$, is 72459 galaxies.  The total sample includes 33951 (47\%) star-forming galaxies. Note that we exclude the redshift range of $0.14<z<0.17$ due to sky line contamination of the H$\alpha$ emission line \citep[e.g.][]{Madusha2011}.  

In addition to star formation, active galactic nuclei (AGN) can also contribute flux to the H$\alpha$ emission line.  We identify all galaxies that show AGN contamination based on available emission line diagnostics, but with our current dataset, it is not possible to completely disentangle the contribution of these two components.  We use the \citet{Kewley2001} prescription to identify the presence of AGN-dominated spectra based on the diagnostic of \citet*{BPT1981} comparing the ratios of $\mathrm{[NII]/H\alpha} $ and  $\mathrm{[OIII]/H\beta}$.    For galaxy spectra that do not have sufficient signal-to-noise in all four required emission lines, we use either of two separate two-line diagnostics to identify contaminants.  If a galaxy has log$( \mathrm{[OIII]/H\beta )}>1$ or log$( \mathrm{[NII]/H\alpha )}>0.2$ then it is classified as an AGN and removed from the sample.   {In this way we identify 2069 galaxies hosting AGN, which is 3\% (2069/72459) of the full sample.

%%%%%%%%%%%%%%%%%%%%%%%%%%%%%%%%%%%%%%%%
%%              SFR             %%
%%%%%%%%%%%%%%%%%%%%%%%%%%%%%%%%%%%%%%%%

\section{Methods}\label{sec:meth}

\subsection{Star Formation Rates}\label{sec:sfr}

%-----  --------------------- FIGURE-----------------------------%
%      3.5 MB
\begin{figure}
\includegraphics[width = 0.5\textwidth]{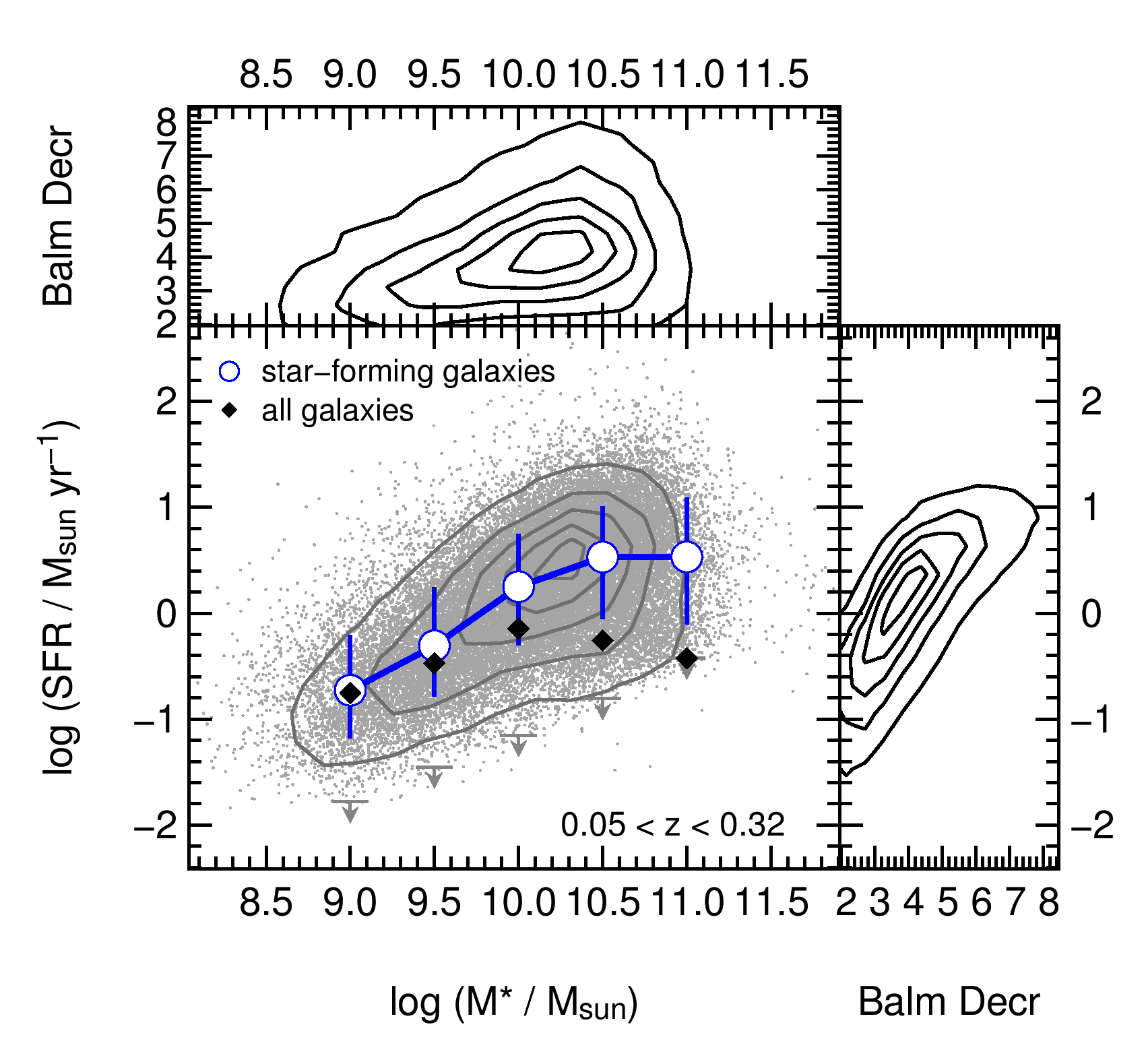}
\caption{\label{fig:sfr_mass}
Star formation rate versus stellar mass for the redshift range of $0.05<z<0.32$.  Light grey points represent individual star-forming galaxies and the contours enclose 10/30/50/70/90\% of these data.  The large open circles are median values of the dust-corrected SFRs in bins of stellar mass for star-forming galaxies.  Since the errors on the median values are very small, we show the spread of values as error bars.  Solid black diamonds show median SFRs for all galaxies in the full redshift range.  The grey arrows show the limits of detectable SFR, calculated assuming the average redshift of galaxies in each stellar mass bin, an H$\alpha$ equivalent width of 3\AA~and the $r$-band limit of the survey.  Also shown are Balmer decrements (BD) as a function of stellar mass and SFR.}
\end{figure}
%-----   --------------------- FIGURE-----------------------------%

We use H$\alpha$ luminosity to determine SFRs for GAMA galaxies.  Once corrected for stellar absorption, the H$\alpha$ equivalent width (EW) can be used, along with an estimate of the continuum luminosity for the galaxy, to recover an effective H$\alpha$ line luminosity for the entire galaxy.  

We attempt to correct for galaxy light lost outside the 2 arcsec diameter optical fiber by applying an aperture correction.  Aperture corrections are based on the absolute magnitude, $M_r$, of each galaxy as an estimate of continuum luminosity, thereby recovering the H$\alpha$ luminosity for the whole galaxy, under the assumption that the flux of the continuum at the wavelength of H$\alpha$ is represented by the flux at the effective wavelength of the r-band filter.  This method of applying aperture corrections to the luminosities is detailed in \citet{Hopkins2003} and \citet{Madusha2011}.  The  aperture correction for GAMA galaxies is typically a factor of 2-4.  In order to minimise the large uncertainties introduced by aperture corrections at the lowest redshifts, we only include galaxies at $z>0.05$ in this study. 

Dust corrections are determined individually for each galaxy by measuring the observed Balmer decrement ($BD$), which is sensitive to the amount of extinction under the assumption of Case B recombination.    We assume the standard theoretical H$\alpha$/H$\beta$ ratio of 2.86, which is valid for ionized gas with an electron temperature $T_e$= 10000~K and electron density of  $n_e=100~cm^{-2}$ \citep{LSanchez2009}.

Stellar absorption corrections are applied to both H$\alpha$ and H$\beta$ fluxes according to Equation 4 in \citet{Hopkins2003}.  Out of all the galaxies with measured Balmer decrements, 9\% have $BD< 2.86$; these are set to 2.86 for the purpose of this investigation.  

Not all galaxies have both H$\alpha$ and H$\beta$ measurements.  For galaxies with only H$\alpha$ measurements, we use the empirical relation between the aperture-corrected luminosity before obscuration correction ($L_{H\alpha,c}$) and the BD determined in \citet{Madusha2013}:

\begin{align} 
BD &=\, 1.003 \, \textrm{log}(L_{H\alpha},c) - 30.0, & \textrm{log}(L_{H\alpha},c) \ge 32.77 \notag \\
&= \,2.86, & \textrm{log}(L_{H\alpha},c) < 32.77
\end{align}

We adopt the method of \citet{Hopkins2003} for calculating total aperture-corrected H$\alpha$ luminosities from fibre spectroscopy \citep[see also][]{Madusha2011,Brough2011}:

\begin{align} 
L_{H\alpha} = &(EW+EW_{c})\,10^{-0.4(M_{r}-34.10)}  \notag \\
 &\times \frac{3\times 10^{18}~[W]}{[6564.61(1+z)]^{2}} \times \left(\frac{F_{H\alpha}/F_{H\beta}}{2.86} \right)^{2.36} 
\end{align} 
where $EW_{c}$ is the constant correction applied to account for stellar absorption in the H$\alpha$ and H$\beta$ Balmer emission lines \citep[$EW_{c} = 2.5$~\AA,][]{Madusha2013,hopkins2013}, $M_{r}$ is the $k$-corrected (to $z=0$) absolute $r$-band Petrosian AB magnitude, and $F_{H\alpha}/F_{H\beta}$ denotes the Balmer decrement used to correct for dust obscuration. 

Finally, the H$\alpha$ SFR is determined from the \citet{Kennicutt1998} relation,  

\begin{equation}
\textrm{SFR}\,[\Msun \,\textrm{yr}^{-1}] = \frac{L_{H\alpha}}{1.27 \times10^{34}}
\end{equation}

\noindent  We adjust the SFR to a \citet{Chabrier2003} initial mass function (IMF) by dividing by a factor of $1.5$.  

For this study, we use galaxies with $r_{AB}<19.4$ because all three GAMA phase 1 fields are complete to this magnitude.  We also define star-forming galaxies to be those with measured $EW_{H\alpha} > 3$\AA ~and $F_{H\alpha}>2.5\times10^{-16}$~ergs/s/cm$^{2}$/\AA, in line with previous studies of \citet{Madusha2011} and \citet{Brough2011}.

\subsection{Stellar Mass Estimates}\label{sec:mass}

Stellar masses are estimated for each galaxy by fitting a grid of synthetic spectra to aperture photometry \citep{Hill2011} in five bands: $ugriz$  \citep{Taylor2011}.  The synthetic stellar population spectral models come from \citet{BC03}, assuming a \citet{Chabrier2003} IMF, a \citet{Calzetti2000} dust obscuration law, and an exponentially declining star formation history.   The stellar masses were determined from the most likely mass-to-light ratio in the $i$-band, over the full range of possibilities provided by the grid.  See \citet{Taylor2011} for the full details of the method. 

%%%%%%%%%%%%%%%%%%%%%%%%%%%%%%%%%%%%%%%%
%%              SFR - stellar mass                %%
%%%%%%%%%%%%%%%%%%%%%%%%%%%%%%%%%%%%%%%%

\section{Results}\label{sec:results}

\subsection{Star Formation Rates and Stellar Mass}\label{sec:sfr-mass}

The relationship between SFR and stellar mass is shown in Figure~\ref{fig:sfr_mass}.  All individual star-forming galaxies are presented as light grey points with contours enclosing 10, 30, 50, 70, and 90\% of the data.  Blue open circles represent the median values of SFR for all star-forming galaxies.  The grey arrows in Figure~\ref{fig:sfr_mass} show the limits of detectable SFR, calculated assuming the average redshift of galaxies in each stellar mass bin, an H$\alpha$ equivalent width of 3\AA~and the $r$-band limit of the survey.    

We see evidence in Figure~\ref{fig:sfr_mass} for SFRs to increase as a function of stellar mass for low-mass galaxies ($M_{*}<3\times10^{10}$ $\Msun$) over the full redshift range of $0.05<z<0.32$.   At high masses, the relationship flattens such that SFRs are consistent with remaining constant for $M_{*}>3\times10^{10}$ $\Msun$.  

%and start to decrease, on average, for the most massive galaxies, with $M_{*}>1\times10^{11}$ $\Msun$. 

%-----  --------------------- FIGURE-----------------------------%
%   3.7 MB
\begin{figure}
\includegraphics[width = 0.5\textwidth]{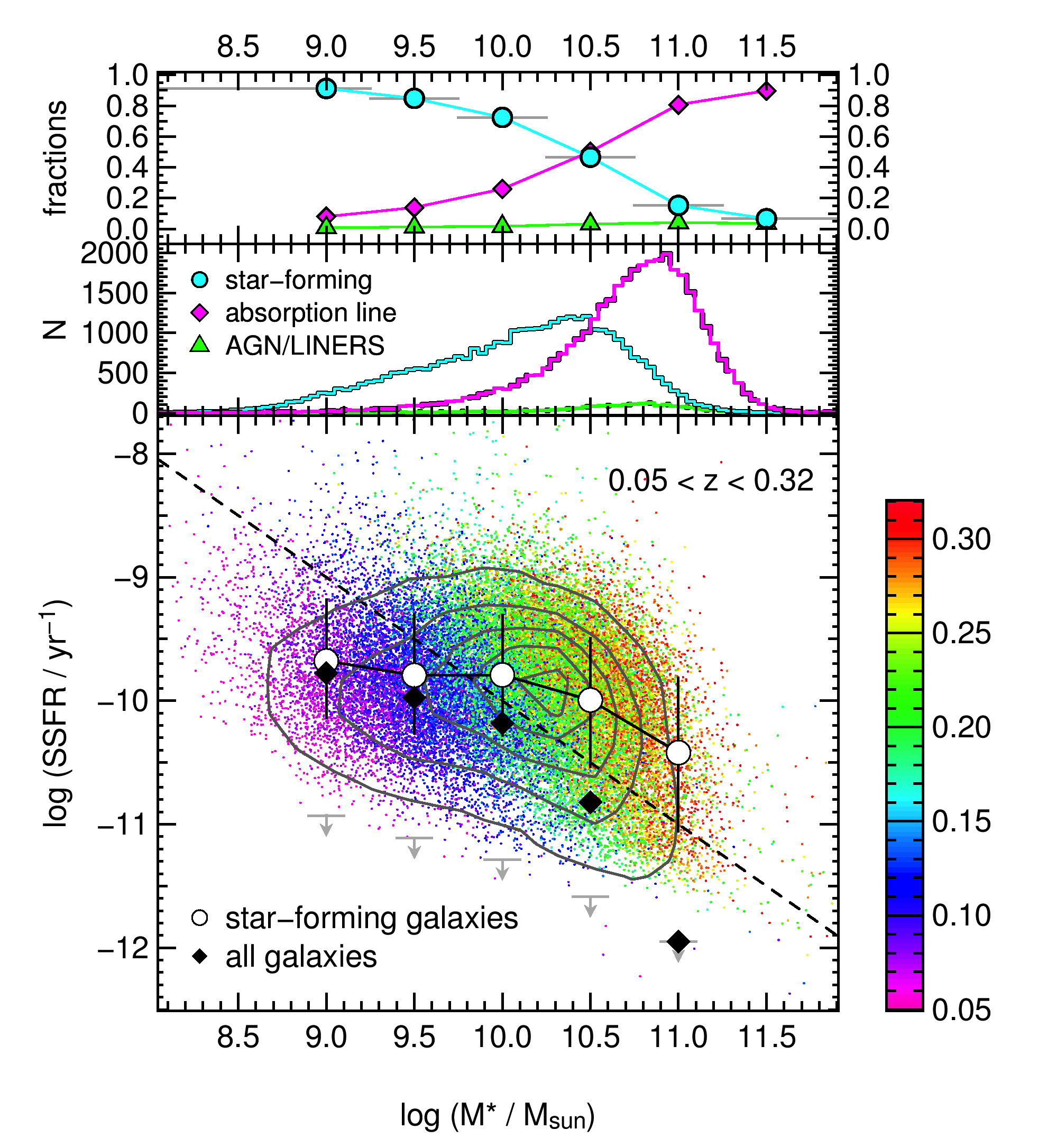}
\caption{\label{fig:ssfr_mass_zcolors}
Specific star formation rate against stellar mass for galaxies of $0.05<z<0.32$ (colours of points show redshift).  The large open circles are median values of SSFR in mass bins for star-forming galaxies only.  Standard deviation errors are shown. Solid diamonds show median values of SSFR for all galaxies.  The dashed line corresponds to SFR = 1  $\Msun$/yr.  The middle panel shows the distribution of star-forming, AGN/LINER (from the BPT diagram or other 2-line diagnostics), and absorption line (no AGN or H$\alpha$ emission) systems.  The top panel shows these categories of galaxies as fractions. }
\end{figure}
%-----   --------------------- FIGURE-----------------------------%

To ensure that this behaviour is not simply due to the higher dust corrections applied on average to galaxies with higher stellar masses or higher intrinsic SFRs, we test the median values of SFR before the dust corrections are applied and find no difference in the slope of the SFR-M$_{*}$ relation for star-forming galaxies.  Shown in Figure~\ref{fig:sfr_mass} are the calculated Balmer Decrements as a function of both stellar mass (top panel) and SFR (right panel).  Dust content is more strongly related to the amount of active star formation than to the stellar mass of the galaxy \citep{Bauer2011_GNS}.  

Galaxies that are not observed as star-forming galaxies based on our criteria do not appear as individual grey points in Figure~\ref{fig:sfr_mass}.  Including these galaxies in the calculations of median SFRs greatly affects the results.  We assign the mass-dependent upper limit values of SFR to all non-star-forming galaxies in the full sample.  We then include these values in calculating median SFRs for all galaxies, shown as the solid black diamonds in Figure~\ref{fig:sfr_mass}.  We find that low-mass galaxies still show a trend for increasing average SFRs with increasing stellar mass, but then at roughly $M_{*}<3\times10^{10}$ $\Msun$, the average SFRs steeply drop to low values, due to the increasing proportion of massive, quiescent systems.    

The consistency between different SFR indicators, in particular for FIR, UV, [OII], and H$\alpha$, is relatively robust for large samples, although it can vary dramatically for individual galaxies \citep[e.g.][]{Hopkins2003,DW2011}.  This broad consistency, demonstrated explicitly within the GAMA sample by \citet{DW2011}, is sufficient that there are unlikely to be any substantial systematic effects introduced by our comparison of the GAMA sample with those at high redshift, which we will begin to explore in Section~\ref{sec:sfh}.

%%%%%%%%%%%%%%%%%%%%%%%%%%%%%%%%
%%              SSFR - M*            %%
%%%%%%%%%%%%%%%%%%%%%%%%%%%%%%%%

\subsection{Specific Star Formation Rates}\label{sec:ssfr-mass}

We show above that the SFR values of massive star-forming galaxies are on average higher than the SFRs of low-mass galaxies.  In order to understand how significant star formation is for the growth and assembly of stellar mass for galaxies of different masses, we look at the SFR per unit stellar mass, or the specific star formation rate (SSFR~=~SFR~/~M$_{*}$).

%-----  --------------------- FIGURE-----------------------------%
% 14 MB
\begin{figure}
\includegraphics[width = 0.52\textwidth]{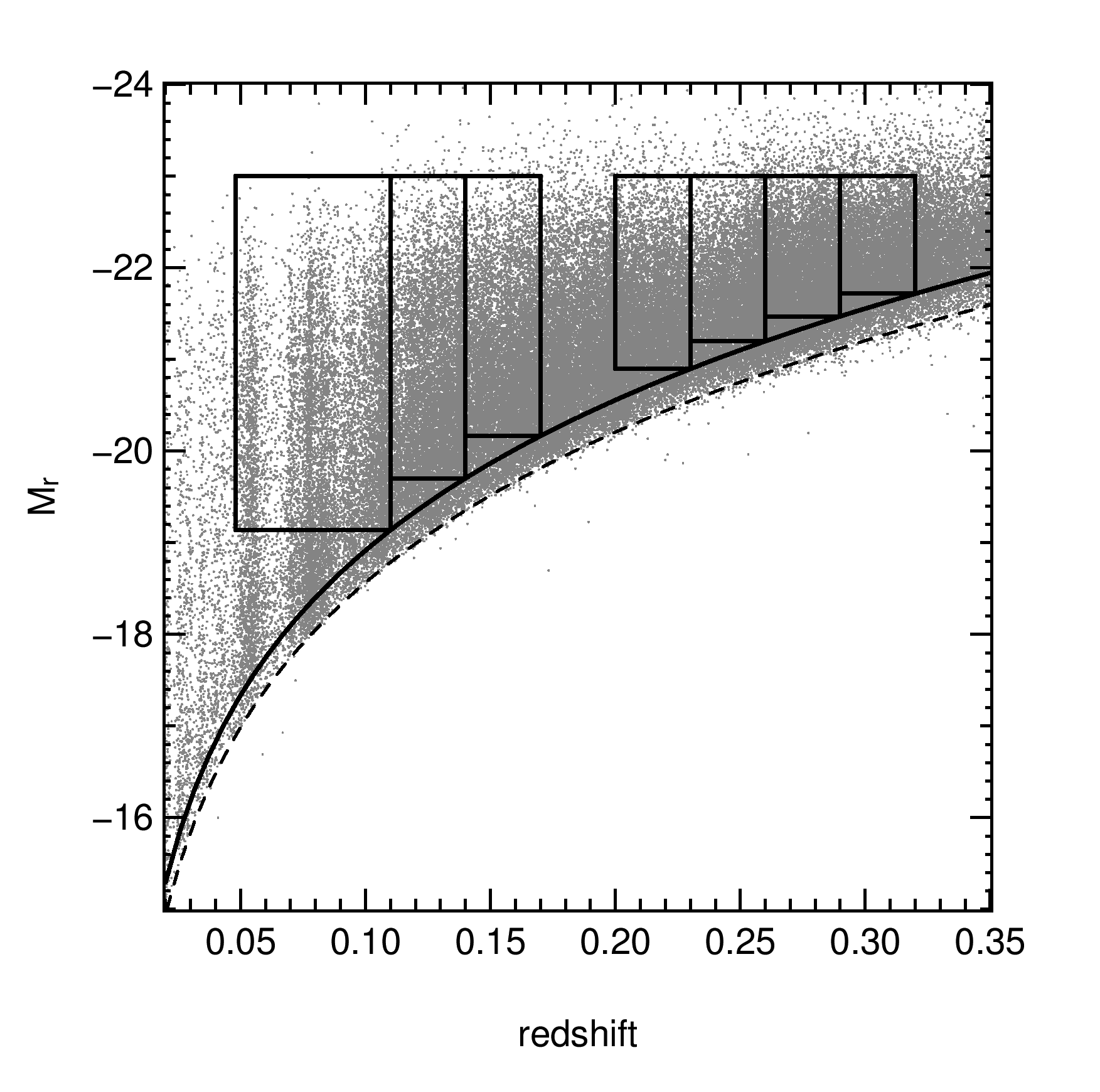}
\caption{\label{fig:gama_abs}
Absolute Petrosian magnitude as a function of redshift for all galaxies within the GAMA volume.  The solid black curve is the $r$-band completeness limit of GAMA of 19.4 mag.  The dashed black line shows the $r_{AB}=19.8$ limit achieved for the 12 hr GAMA field.   The boxes are the redshift-dependent magnitude cuts for each redshift bin.  }
\end{figure}
%-----   --------------------- FIGURE-----------------------------%

Figure~\ref{fig:ssfr_mass_zcolors} shows SSFR as a function of stellar mass over the full redshift range of $0.05<z<0.32$.  Median values are shown as open circles for the full dust-corrected sample of star-forming galaxies.  The median SSFRs decrease with increasing stellar mass over the full stellar mass range, independent of the dust correction.  The SSFR-M$_{*}$ relation is not a constant function of stellar mass, and shows a drop in SSFRs for high mass galaxies.  There is also a hint of an upturn in the average SSFR for the lowest stellar mass galaxies.  

As galaxies stop forming stars, their SSFRs decrease until they fall below our detection limits and do not appear as individual points in Figure~\ref{fig:ssfr_mass_zcolors}.  The grey arrows in Figure~\ref{fig:ssfr_mass_zcolors} show our detection limits, which are calculated assuming an H$\alpha$ equivalent width of 3\AA~(prior to the stellar absorption correction) and the $r$~band magnitude limit of the survey at the median redshift of the galaxies in each stellar mass bin.    We assign these mass-dependent values of SSFR to all non-star-forming galaxies in each stellar mass bin, then include these in the median SSFRs, shown as the solid black diamonds in Figure~\ref{fig:ssfr_mass_zcolors}.  The median SSFRs for the full sample show a very strong decline with increasing stellar mass.

In order to quantify the fraction of galaxies not forming stars as a function of stellar mass, we show the star-forming fraction in the top panel of Figure~\ref{fig:ssfr_mass_zcolors}.  We find that 70-80\% of $M_{*}<10^{10}$ $\Msun$ galaxies are forming stars.  This fraction steadily decreases with increasing stellar mass as higher mass systems become dominated by absorption-line spectra and AGN activity.  We find that for galaxies with $M_{*}>10^{11}$ $\Msun$, only 10\% are forming stars, with an average $\log(\rm{SSFR/yr^{-1}})=-10.5$.

The middle panel of Figure~\ref{fig:ssfr_mass_zcolors} shows the observed stellar mass distribution of galaxies from the full sample that are identified as star-forming (cyan), AGN (green), and otherwise (magenta).  The distribution of stellar masses for non-star-forming galaxies and AGNs covers a similar stellar mass range, peaking at $M_{*}\sim8\times10^{10}$ $\Msun$.  The distribution of stellar masses for star-forming galaxies peaks at a lower stellar mass of $M_{*}\sim5\times10^{10}$ $\Msun$ and extends down to the lowest masses probed.

%-----  --------------------- FIGURE-----------------------------%
% 2.7 MB
\begin{figure*}
\includegraphics[scale=0.9]{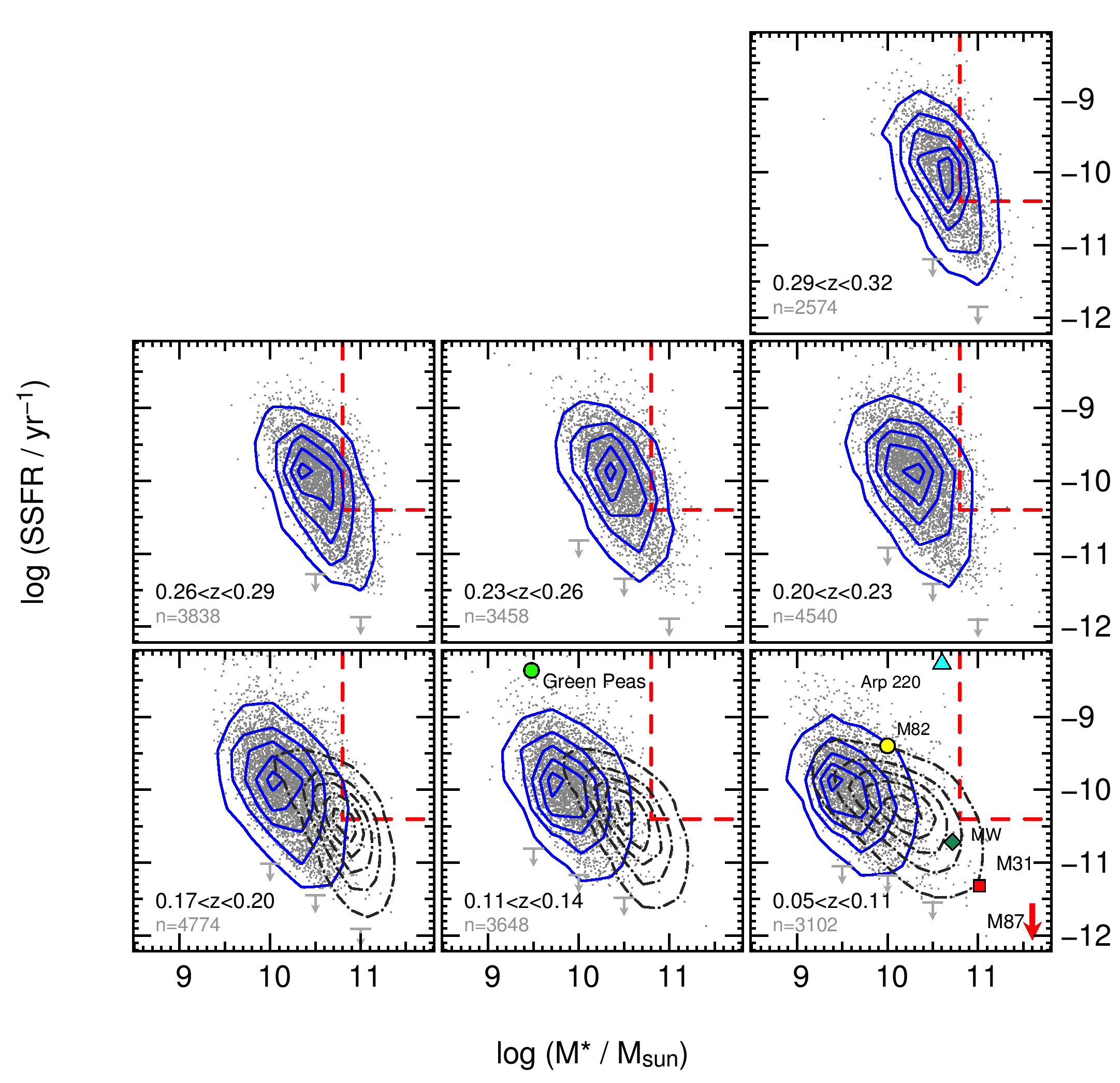}
\caption{\label{fig:gama_ssfr}
SSFR versus stellar mass in redshift bins for GAMA (blue contours) and the full SDSS-DR7 survey (dark grey dash-dot contours).  Light grey points represent individual star-forming GAMA galaxies.  The red dashed lines are identical in each redshift bin and are there to guide the eye.  The bottom left of each panel shows the redshift range and number of GAMA galaxies included in each bin.  In the lowest redshift bin (bottom right), individual galaxies are represented:  Arp 220, M82, the Milky Way, M31, M87, covering a wide range of SSFR.  In the bottom left panel, we show the location of the average population of ``Green Pea'' galaxies.   
 }
\end{figure*}
%-----   --------------------- FIGURE-----------------------------%

One complication with this general representation of the data is that it is difficult to ascertain the evolution of stellar mass growth in these galaxies as a function of stellar mass because both redshift dependence and stellar mass dependence contribute to the trends seen in Figure~\ref{fig:ssfr_mass_zcolors}.  This is demonstrated by the colours of the points representing redshifts of individual galaxies.  Lower mass galaxies are only detected at the low redshift end of the range, as they become too faint to be detected at higher redshifts.  

In order to distinguish the trends with stellar mass from those with redshift, we split the sample into seven redshift bins.  We use volume-limited redshift bins based on cuts applied in absolute Petrosian magnitude as a function of redshift for all galaxies within the GAMA volume, shown in Figure~\ref{fig:gama_abs}.  Grey points show all galaxies from the GAMA survey in addition to all SDSS galaxies within the GAMA survey area on the sky.  The dashed curve in Figure~\ref{fig:gama_abs} shows the $r$-band completeness limit of GAMA of 19.8 mag.   The solid black curve, used for this work, shows the $r=19.4$ completeness limit achieved for all three GAMA fields.  The boxes show the redshift-dependent absolute magnitude cuts for each redshift bin.  In order not to introduce measurement-based biases, we exclude the galaxies between $0.14<z<0.17$, as we have done throughout the paper so far.  This range is excluded as it encompasses the narrow redshift range where sky lines interfere with the measured H$\alpha$ line \citep[see][]{Madusha2011}.  For the results presented throughout the rest of this manuscript, we use only the sample of galaxies in these volume-limited redshift bins.

%%%%%%%%%%%%%%%%%%%%%%%%%%%%%%%%%%%%%%%%
%%              SSFR - stellar mass      -- zed bins          %%
%%%%%%%%%%%%%%%%%%%%%%%%%%%%%%%%%%%%%%%%

\subsection{SSFRs, stellar mass, and redshift}\label{sec:ssfr-mass-zed}

We show the SSFR versus stellar mass split into 7 volume-limited redshift bins in Figure~\ref{fig:gama_ssfr}.  Decreasing in redshift from $z\sim0.32$ at upper right to $z\sim0.05$ at bottom right, we show all galaxies within the GAMA fields as grey points in each bin, and as blue contours that encompass 10\%, 30\%, 50\%, 70\%, and 90\% of the galaxies, normalised by the volume in the redshift bin.   The vertical dash-dot grey lines show approximate stellar mass limits in each redshift bin \citep{Taylor2011} and the arrows show SSFR limits in stellar mass bins at each redshift.    

It is clear from Figure~\ref{fig:gama_ssfr} that the relationship between SSFR and stellar mass is not flat at any of these redshifts, rather the SSFR declines rapidly with increasing stellar mass.  The lowest-mass galaxies exhibit the highest SSFRs at all redshifts.  There exists an upper envelope in the SSFR-M$_{*}$ plane that increases with increasing redshift such that high-mass galaxies at $z>0.3$ show much higher SSFRs than high-mass galaxies at the lowest redshifts.  The red dashed lines are identical in each redshift bin, in order to highlight this effect.  In the highest redshift bin, at the upper right of Figure~\ref{fig:gama_ssfr}, there is a large population of high-mass galaxies with high SSFRs inside the region marked by the red dashed lines.   The number of galaxies occupying this space decreases with decreasing redshift such that very few galaxies occupy this region of the SSFR-M$_{*}$ plane by $z<0.1$.  

In order to investigate whether this effect is due to the decreasing volume probed as redshift decreases, we show the full SDSS Data Release 7 catalog \citep{DR72009} as dark grey dash-dot contours in each redshift bin up to $z=0.2$.  The 90\% inclusive contours for the SDSS data encompass galaxies of slightly higher stellar masses than GAMA, as expected due to the larger volume observed, but we still find a distinct decrease in the population of SDSS galaxies within the red box as redshift decreases.  The centres of the distributions are offset due to GAMA spectroscopy reaching nearly two magnitudes fainter than SDSS.  The conclusion is that the high mass star-forming galaxies, which make up $\sim$~20\% of all massive galaxies at these redshifts, are rapidly shutting down star formation over the $\sim$3.5 Gyr timeframe since $z=0.32$  shown in Figure~\ref{fig:gama_ssfr}. 

Even as high mass galaxies shut down star formation and the upper envelope of SSFR steadily decreases, we find that at every redshift to $z=0.32$, the lowest stellar mass galaxies detectable show very high SSFRs; higher than expected from simple models of exponentially declining star formation, as we will discuss in Section~\ref{sec:sfh}.  Certainly low-mass galaxies with both low and high SSFRs exist that do not appear in our sample. The former is due simply to the spectroscopic detection limit for H$\alpha$, and is indicated by the detection limits shown in Figure~\ref{fig:ssfr_mass_zcolors}. The latter is a consequence of galaxies that may have $r$-band magnitudes below the GAMA survey limit, regardless of their brightness in H$\alpha$. This bivariate selection effect is discussed at length in \citet{Madusha2013}. We emphasise that our analysis throughout is based on the existence of the detected population of high SSFR low-mass galaxies, and our conclusions are not influenced by the selection effects preventing the inclusion of the low SSFR low-mass galaxy population.

In order to begin to demonstrate the significance of the high SSFRs of low-mass galaxies, we show the location of well-studied local galaxies in the lowest redshift panel of Figure~\ref{fig:gama_ssfr}.   The galaxies shown include the Milky Way, our local group neighbour, M31, the star-forming galaxy, M82, the interacting system classified as an ultra luminous infrared galaxy, Arp220, and the large elliptical galaxy at the centre of the Virgo cluster, M87, which falls far below the detection limits of this study since it is not an actively star-forming galaxy.  Although extreme starbursts like Arp220 are the most common mode of star formation in galaxies at $z\sim2$  \citep{Daddi2007,Bauer2011_GNS}, there are no galaxies with similar stellar mass harbouring such high SSFRs in the GAMA volume at low redshifts.  Individual galaxies are no longer forming stars with such burst strengths by $z<0.32$.  

An example of a population of the high SSFR, low-mass galaxies that we detect in GAMA is the ``green pea'' population of compact, emission line galaxies identified by the Galaxy Zoo project.  These are shown  at their average redshift of $z\sim0.12$ with typical SSFR and stellar mass \citep{Peas} in Figure~\ref{fig:gama_ssfr}.  The ``green peas'' fall just at the upper bounds of observed GAMA sources at their average redshift.

%%%%%%%%%%-------------------------------------------
\subsection{Star-Forming Galaxies versus the Full Sample}\label{sec:all}

%-----  --------------------- FIGURE-----------------------------%
% 2.8 MB
\begin{figure*}
\includegraphics[scale=0.85]{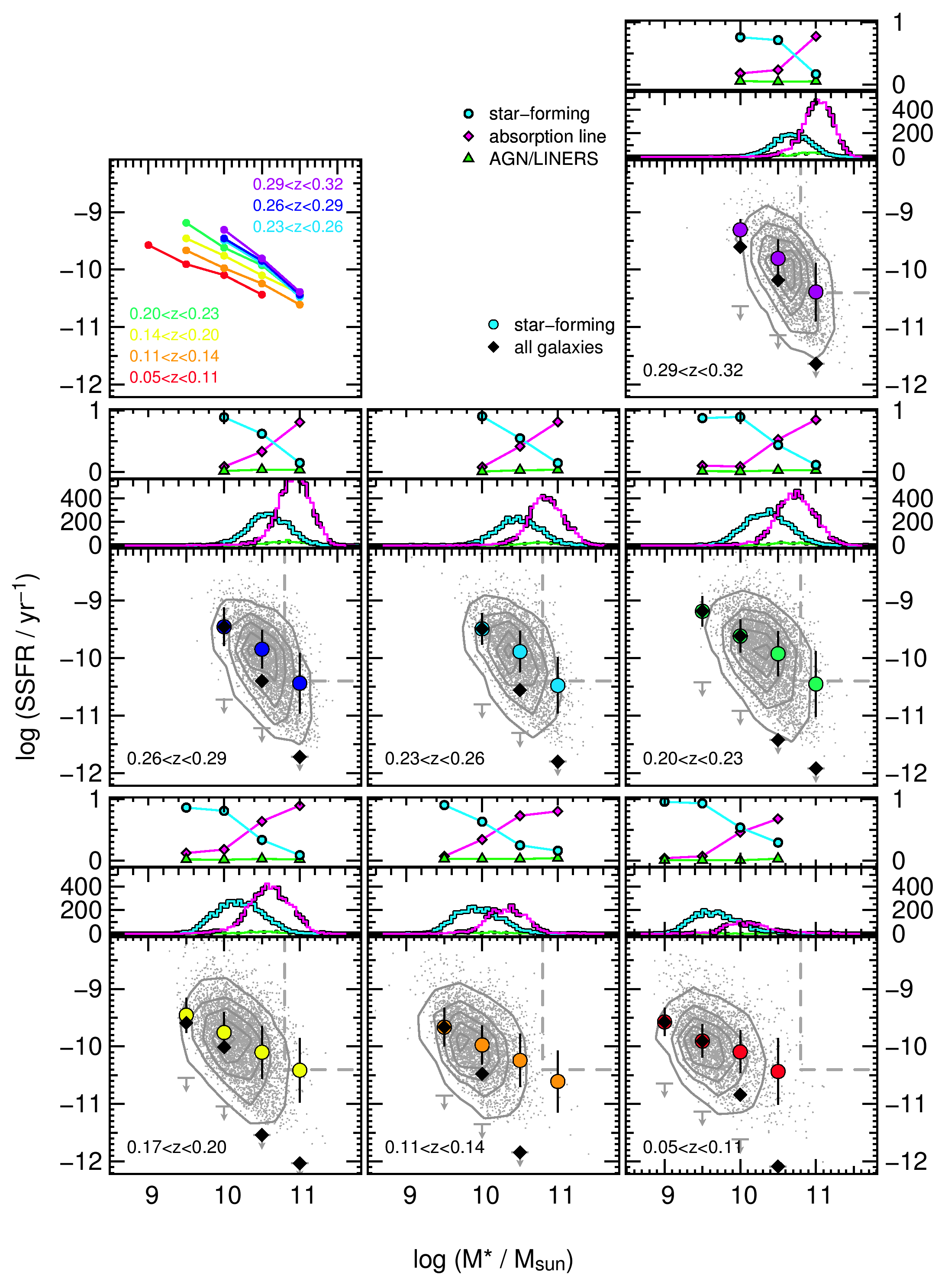}
\caption{\label{fig:ssfr_27}
SSFR vs stellar mass in redshift bins for GAMA as in Figure~\ref{fig:gama_ssfr}.  Light grey points represent individual star-forming galaxies and the contours enclose 10/30/50/70/90\% of these data.  Large circles and solid diamonds show median SSFRs in bins of stellar mass for star-forming and the full sample of galaxies, respectively.  Also shown are the stellar mass distributions of galaxies in each redshift bin and fractions of star-forming (cyan), AGN/LINERS (green), and absorption line (magenta) galaxies.  The top left panel shows the median relations for the star-forming galaxies in each of the redshift bins.  The colours correspond to those shown in each individual redshift bin.
 }
\end{figure*}
%-----   --------------------- FIGURE-----------------------------%

In the previous section we identify a population of low-mass star-forming galaxies with higher SSFRs than high-mass star-forming galaxies, and that the upper envelope decreases such that high mass galaxies have significantly lower SSFRs at every redshift over the $\sim$4 Gyr period covered in Figure~\ref{fig:gama_ssfr}.  In this section we begin to look at properties of the entire sample of galaxies, including those not forming stars and those identified as LINERS or AGN.  

Figure~\ref{fig:ssfr_27} includes all the information presented in Figure~\ref{fig:gama_ssfr} in grey, with the addition of median SSFRs in stellar mass bins for star-forming galaxies (circles) and the full sample (black diamonds).  Note that the ``full sample'' includes star-forming and absorption line galaxies, but excludes those identified as AGN or LINERS via the BPT diagram (see Section~\ref{sec:data}).  AGNs and LINERS make up $\sim$3\% of the population over the full redshift range and broadly distributed across the stellar mass range, as seen in Figure~\ref{fig:ssfr_27}.  

The relationship between SSFR and stellar mass is not flat at any redshift up to $z=0.32$.  SSFR decreases with increasing stellar mass for both the star-forming and full sample of galaxies.  The relationship between SSFR and stellar mass is steeper for the full sample of galaxies than it is for just the star-forming population.  Star-forming galaxies dominate the population of M$_{*}~<~10^{10.5}$ $\Msun$ galaxies.   Care should be taken when deriving and interpreting relationships between SFRs, SSFRs and stellar mass, since the inclusion or not of the quiescent population greatly affects the result.  

It is interesting to note that we do see evolution in the median values of SSFRs in the redshift range studied here, for the detected star forming population.  The top left panel of Figure~\ref{fig:ssfr_27} shows the median values of SSFR versus stellar mass for all redshift bins, for the star formers. We caution that this result may be influenced by our H$\alpha$ detection limits, as discussed in \S\,\ref{sec:ssfr-mass-zed}. If a significant population of star forming galaxies at any given mass lies below our detection limit, the median would be lower, reducing the observed effect. The
extent of this impact can be gauged by the evolution seen in the galaxy population as a whole, where quiescent systems are assigned to the detection limits for estimating the median SSFR. The fact that the median values of SSFR for the full population at the lower masses are essentially identical to that from the star forming population, a consequence of the low-mass systems being dominated by the star forming population, suggests this is likely to be a small effect at the low-mass end. The slope of the relationship does not change significantly with time, indicating that the dependence of SSFR on stellar mass since $z=0.32$ remains largely unchanged. Again, the slope may steepen to higher redshifts if a significant non-detected population of star formers exists at the higher mass end. For a given stellar mass, the median SSFR decreases by about 0.5\,dex over this redshift range, although the change is smaller for galaxies above M$_*$ = $3\times 10^{10}$ $\Msun$.

It is also interesting to note that the stellar mass at which the proportion of quiescent systems approximately equals that of star formers increases with redshift, from M$_{*} = 10^{10}$ $\Msun$ at the lowest redshifts, to M$_{*} \approx 10^{10.7}$ $\Msun$ in our highest redshift bin.

%%%%%%%%%%%%%%%%%%%%%%%%%%%%%%%%%%%%%%%%
%%              SFH - tau models                %%
%%%%%%%%%%%%%%%%%%%%%%%%%%%%%%%%%%%%%%%%

\section{Star Formation Histories}\label{sec:sfh}

Having presented the median values of SSFRs with stellar mass since $z=0.32$ for the star-forming and full galaxy samples, we can now investigate how galaxies might evolve over the last 3.5 billion years, the timeframe covered by the GAMA sample from $z=0.32$ to $z=0.05$.   We first consider a scenario where galaxies undergo a constant SFR over time, and then examine a mass-dependent, exponentially declining star formation history (SFH), with parametrizations derived from the GAMA sample and those predicted from $z\sim1$ samples.

We choose the latter scenario as it is the simplest already demonstrated to be consistent with galaxy population properties \citep[e.g.][]{Noeske07_tau}. Of course we are only testing, at a given redshift, whether the galaxies observed at that redshift are consistent or not with the model parameters established from the higher redshift population, not the actual star formation histories of any individual galaxy. The goal is to test whether such a simple model is sufficient or not to describe the properties of galaxies observed at any given redshift.
 
\subsection{Constant Star Formation Histories}\label{sec:constant_sfh}

First, consider as an example a relatively low-mass galaxy of M$_*$ = $10^{9}$ $\Msun$, at $z=0.6$, forming stars at a constant observed rate of SFR  $= 10~\Msun$ yr$^{-1}$.  If the galaxy were to continue forming stars at this rate over the period from $z=0.6$ to $z = 0.1$ (roughly 3 Gyr), its mass would increase by a factor of thirty by $z=0.1$.  The galaxy would shift down and to the right in the SSFR-M$_{*}$ plane of Figures~\ref{fig:gama_ssfr}~and~\ref{fig:ssfr_27}, and would stop after increasing by 1.5 dex in stellar mass.  A constant  SFR  $= 1~\Msun$ yr$^{-1}$ over the same time period would instead shift the galaxy by 0.3 dex in stellar mass.  A constant SFR over a long period, however, at least for higher mass galaxies, is known not to be a good model, as it implies too high a stellar mass in the local universe \citep{Bell2007,Sharp2010} and does not match the SFRs of locally observed galaxies.

\subsection{Exponentially Declining SFHs}\label{sec:exp_sfh}

Since constant star formation histories are not sufficient for building present day observed galaxies, we follow other authors in considering a closed-box, simple model parametrization of the SFH as an exponentially declining SFR with mass-dependent values of formation redshift and star formation timescale.  Such mass-dependent models were first introduced several decades ago by \citet{Tinsley1968}, have more recently been developed by \citet{Noeske07_tau}, and are successfully employed in other studies as well \citep[e.g.][]{Martin07, Gilbank2011}.

We model the star formation rate as       

\begin{equation} \label{eq:sfr_exp}
\textrm{SFR}(M,t) = \frac{M_{g}}{\tau (1-R)} e^{-t/\tau},
\end{equation}
where $\tau$ is the star formation time scale ($e$-folding time) and $t=t(z) - t(z_{f})$ is the difference between the cosmic time at which the galaxy is observed, $t(z)$, and the time at which the galaxy formed, $t(z_{f})$, and M$_{g}$ is the initial gas mass available at $t(z_{f})$.  The recycled gas fraction, $R$, is set to a value of 0.5 (Bell et al. 2003; Villar et al. 2011).  The stellar mass is then calculated by taking the integral of this SFR from $t(z_{f})$ to the time of observation, which gives

\begin{equation} \label{eq:mass}
M_{*}(M,t) = M_{g} [1-(e^{-t/\tau})].
\end{equation}
Combining these equations, we find that the normalised SSFR evolves with time as 

\begin{equation} \label{eq:ssfr_exp}
\textrm{SSFR}(M,t) = \frac{1}{\tau \,[e^{t/\tau} - 1]},
\end{equation}
where we define the variables $\tau$ and the formation redshifts, $z_f$, to be functions of stellar mass as

%%%%%%     tau    %%%%%%%%%%%%%%%
\begin{equation} \label{eq:tau_exp_test}
\tau = c_{\alpha}\, \mathrm{M_*}^{\alpha},
\end{equation}
%%%%%%    z form     %%%%%%%%%%%%%
\begin{equation} \label{eq:z_f} 
z_{f} = c_{\beta}\,  \mathrm{M_*}^{\beta}.
\end{equation}
 %%%%%%%%%%   end  equation   %%%%%%%%%%
We vary the parameters, $\alpha$ and $\beta$, and the normalisation constants, $c_{\alpha}$ and $c_{\beta}$ to find the best representation of the average SSFR in each redshift bin simultaneously, and provide a unique solution that links stellar mass, formation time, and star formation history.

%-----  --------------------- FIGURE-----------------------------%
% 2.8 MB
\begin{figure*}
\includegraphics[scale=0.9]{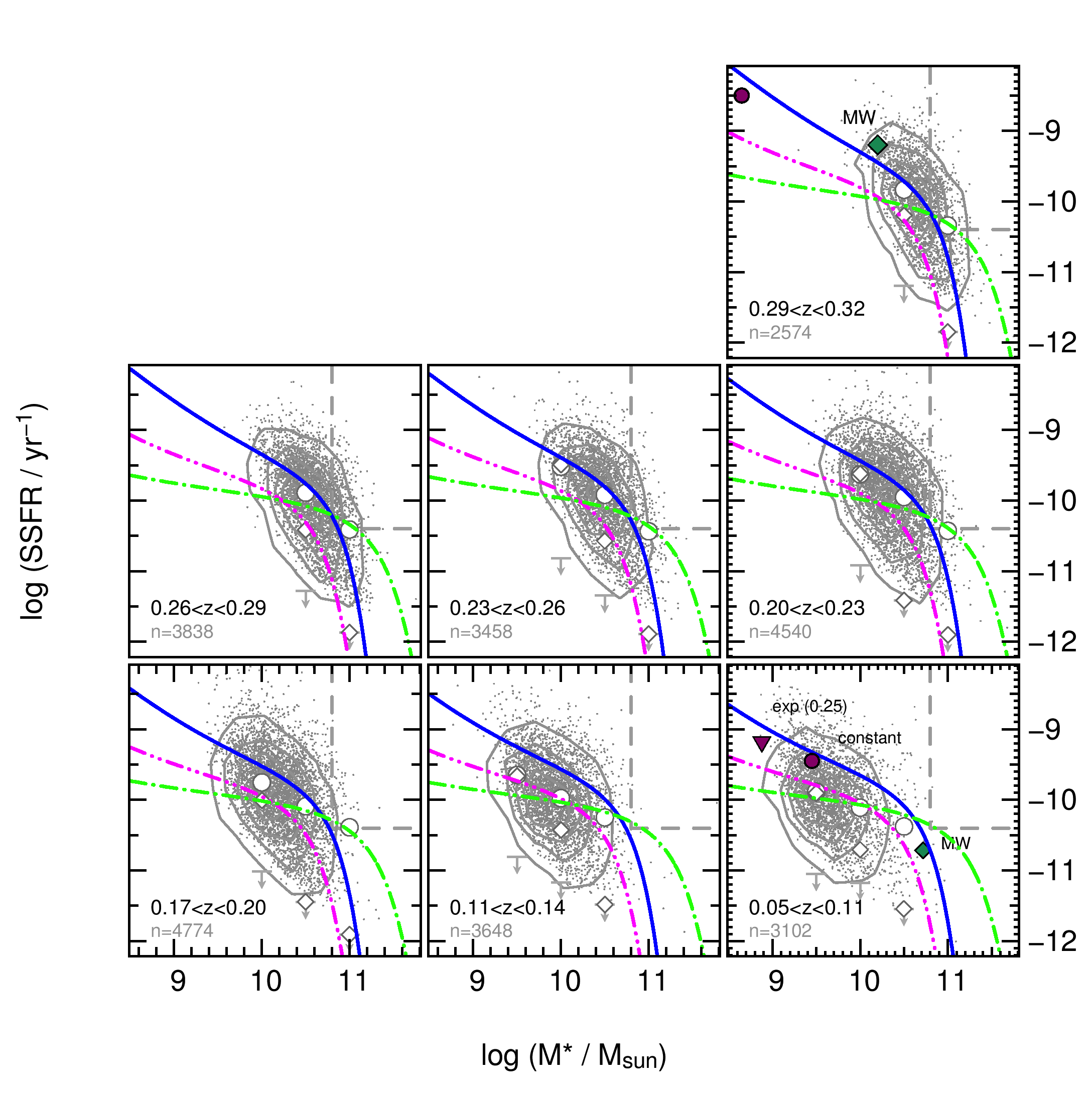}
\caption{\label{fig:gama_ssfr_manytracks}
SSFR vs stellar mass in redshift bins for GAMA as in Figures~\ref{fig:gama_ssfr}~and~\ref{fig:ssfr_27}.  Two parametrizations of exponentially declining star formation histories (SFH) derived from the GAMA sample are shown as the solid blue and dash-dot-dot magenta curves (see Section~\ref{sec:gama-sfh}).  Gilbank et al. (2011) SFHs derived from a $z\sim1$ sample are shown as green dash-dot curves. Magenta shapes show possible evolution of galaxies with constant star formation and a duty cycle of 25\%, as described in Section~\ref{sec:lowmass-sfh}.}
\end{figure*}
%-----   --------------------- FIGURE-----------------------------%

\subsection{SFH Predictions from $z\sim1$}\label{sec:exp_sfh_z1}

One question we can ask is how well SFHs defined to match galaxy populations at $z\sim1$ predict the SSFR distributions measured for the GAMA star-forming galaxy sample.  Figure~\ref{fig:gama_ssfr_manytracks} is identical to Figure~\ref{fig:gama_ssfr}, with the addition of several SFH curves, as described in Section~\ref{sec:exp_sfh}.  The green dash-dot curves in Figure~\ref{fig:gama_ssfr_manytracks} show the SFH presented in \citet{Gilbank2011} as a best fit to the $z\sim1$ galaxy population, and then evolved to the redshift range covered by the GAMA survey.  The parameters found by \citet{Gilbank2011} ($\alpha=-0.99$, $log(c_{\alpha})=20.42$, $\beta=0.31$, $log(c_{\beta})=-2.68$)   are very close to the parametrizations found by \citet{Noeske07_tau} ($\alpha=-1.0$, $log(c_{\alpha})=20.7$, $\beta=0.3$, $log(c_{\beta})=-2.7$).  

Overall, the tracks originally fit to high redshift galaxies do a reasonable job of matching the intermediate-mass GAMA data, but does not recover the observed distribution of SSFRs of GAMA galaxies over the full stellar mass and redshift range.  The major discrepancy can be seen at the low-mass end in each redshift panel.  The observed low-mass GAMA galaxies include a population with much higher SSFRs than those predicted by the models.  As described in Section~\ref{sec:ssfr-mass-zed}, there are low-mass galaxies below our detection limits, but in this exercise we are examining whether mass-dependent, exponentially declining SFHs can explain the existence of the detected population of high SSFR low-mass galaxies observed.  Figure~\ref{fig:gama_ssfr_manytracks} shows that the model parameters fit to (an incomplete sample of) $z\sim1$ galaxies cannot.  

Despite a reasonable agreement above M$_{*}~>~10^{10.5}$ $\Msun$ for most of the redshift range, below $z\sim0.1$, the $z\sim1$ models predict higher levels of star formation in high-mass star-forming galaxies than actually observed.   This characteristic of the model at high redshift is also seen in Figure 3 of \citet{Gilbank2011}.   The high redshift SFHs over-predict by a large factor the SSFRs compared to the observed values for the full sample of GAMA galaxies at all redshifts.

\subsection{GAMA Star Formation Histories}\label{sec:gama-sfh}

We now present two possible parametrizations of the exponentially declining SFH that cover the range of SSFRs observed for GAMA galaxies between $0.05<z<0.32$.  We attempt to simultaneously reproduce high SSFRs in low-mass galaxies and low SSFRs in high-mass star-forming galaxies.  These models are shown in Figure~\ref{fig:gama_ssfr_manytracks} as the blue solid curves ($\alpha=-1.06$, $log(c_{\alpha})=20.42$, $\beta=0.28$, $log(c_{\beta})=-2.68$) and magenta dash-dot-dot curves ($\alpha=-1.03$, $log(c_{\alpha})=20.0$, $\beta=0.28$, $log(c_{\beta})=-2.5$).  

We find that a single track can reproduce the high SSFRs for low-mass galaxies and low SSFRs in high-mass star-forming galaxies, but not across the full 3.5 Gyr time frame shown in Figure~\ref{fig:gama_ssfr_manytracks}.    We found the most success with this SFH in achieving high SSFRs for low-mass galaxies, matching the median values for star-forming galaxies at all redshifts in the range $0.14<z<0.32$.  At $z<0.14$, this SFH shows higher SSFRs for low-mass galaxies than the calculated medians, but does better than the other models at reproducing the large populations of low-mass galaxies with high SSFR. 

The dash-dot-dot magenta line matches the lowest redshift star-forming galaxies well but underestimates the observed SSFRs for all redshifts above $z=0.14$.  The dash-dot-dot magenta curves do well representing the SSFRs for the full sample and not just the star-forming galaxies.    

These two parametrizations differ mostly in normalisation, not the values of the exponents, and successfully encompass the range of values for the GAMA data.  The results show that the timescales over which galaxies form stars are mass-dependent such that observed high-mass galaxies form stars over short periods of time (from 1 to less than 0.5 Gyr).  Massive galaxies form very early using these parametrizations, as early as $z=9$ for the highest stellar masses.   In the models shown in Figure~\ref{fig:gama_ssfr_manytracks}, galaxies with M$_*$ = $10^{9}$ $\Msun$ could form stars on timescales as long as 2.5 to 3 Gyr, consistent with those found previously by \citet{Thomas2005}.  

In order to reproduce the high SSFR upturn found in at least some low-mass GAMA galaxies with the exponentially declining SFHs, much later formation redshifts are required for low-mass galaxies.  As pointed out in \citet{Gilbank2011}, when the lower stellar mass limit of a sample is quite high, the high values of SSFR and this upturn in SFH models for low-mass galaxies will not be seen.  Once star formation begins, stellar mass grows very quickly through star formation and low-mass galaxies form enough stars, according to the exponentially declining SFH, that these galaxies are no longer low-mass.  

A conclusion from this exercise is that galaxy properties should not be compared simply within the same mass range at different redshifts, even for $0.05<z<0.32$ which is often considered the ``local universe'', because a substantial amount of stellar mass builds up over time through star formation alone, without considering any contribution from mergers.  In order to carefully explore stellar mass growth, samples need to be chosen that encompass the appropriate progenitor mass range at successively higher redshifts.  Galaxy samples could ideally be corrected for this mass growth, which can be done using these models \citep{Noeske2009} for intermediate mass galaxies, as long as the correct parameters can be determined.  To demonstrate this, we can use the example of the Milky Way.

\subsection{The Milky Way at $z=0.32$}\label{sec:MW-sfh}

Using the best parametrizations of the simple exponentially declining SFH found for the GAMA sample (the blue solid line in Figure~\ref{fig:gama_ssfr_manytracks}), we can determine whether the GAMA survey would have detected our Milky Way Galaxy at $z\sim0.32$.  As shown in the highest redshift panel at the top-left of Figure~\ref{fig:gama_ssfr}, the Milky Way would be at the edge of the low-mass, high-SSFR population of galaxies we observe at that redshift.  

The fact that the Milky Way would barely have been detected at $z=0.32$ from the GAMA selection criteria raises some interesting issues.  First, it means that mass-dependent exponentially declining SFHs work sufficiently well at accounting for intermediate- to high-mass present day galaxies like the Milky Way \citep{Thomas2005,Governato2007,Noeske2009,Thomas2011,Gilbank2011}, but it does not account for the bulk of low-mass galaxies.  This is because if the lowest mass galaxies detectable at $z=0.32$ evolve with exponentially declining SFHs to become Milky Way-like galaxies today, then it is difficult to explain the large numbers of low-mass galaxies with high SSFRs at any of these redshifts.  If low-mass galaxies continue to support such SFHs over this time period, they build up stellar mass quickly and rapidly shift to the high mass region in each bin of Figures~\ref{fig:gama_ssfr}~and~\ref{fig:gama_ssfr_manytracks}.  

According to the mass-dependent exponentially declining SFHs, low-mass galaxies form at later times with long star formation timescales.  Yet we observe a large population of positive outliers in Figure~\ref{fig:gama_ssfr_manytracks}: low-mass galaxies with SSFRs well above the values predicted by the models.  Even considering the fact that each redshift bin is not complete in stellar mass and that there are populations of low-mass galaxies with low SSFRs at all of these redshifts, the existence of so many low mass galaxies with high SSFRs warrants further investigation.  

\subsection{Accounting for Low-Mass High-SSFR galaxies}\label{sec:lowmass-sfh}

%-----  --------------------- FIGURE-----------------------------%
\begin{figure}
\includegraphics[scale=0.7]{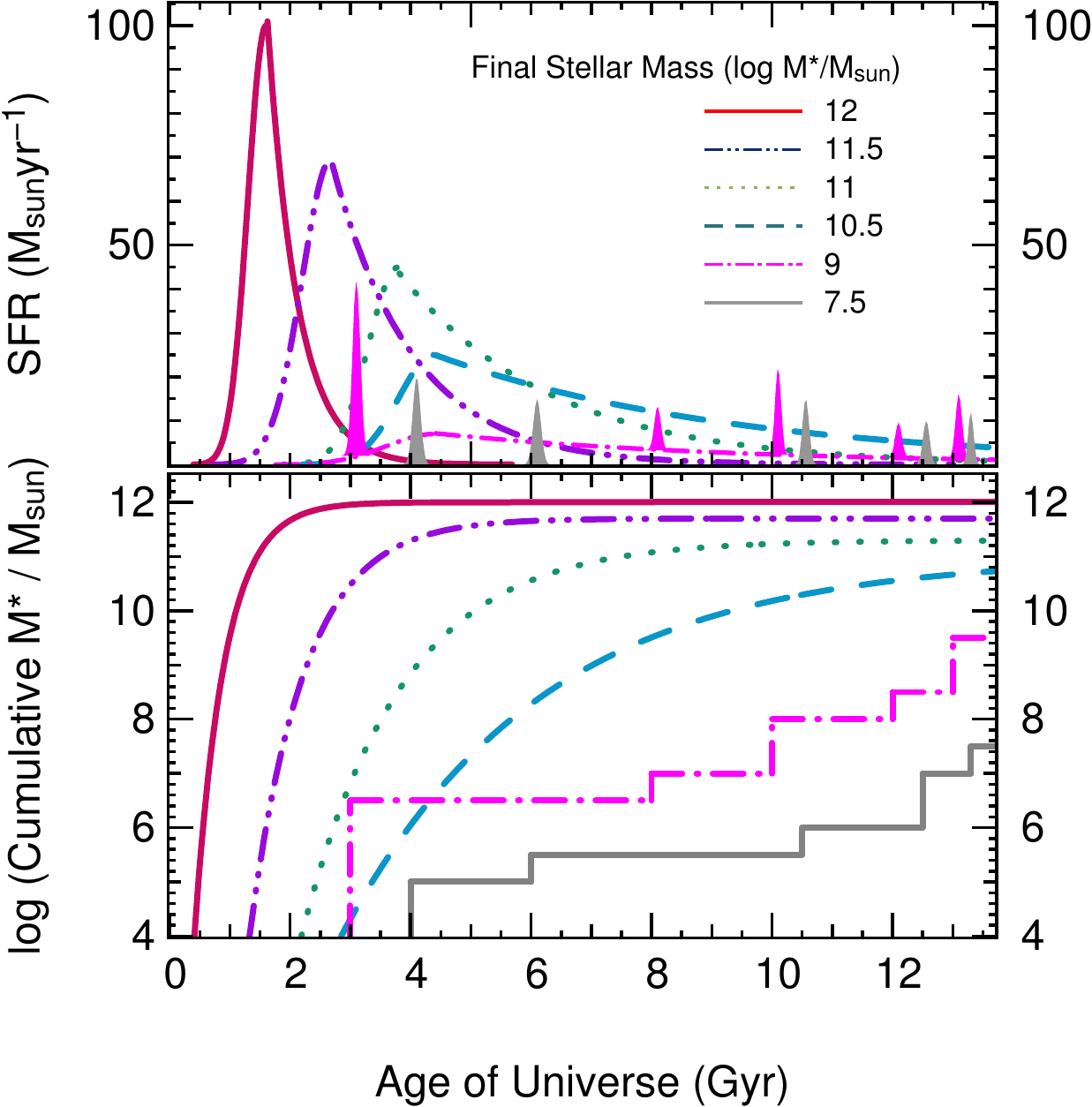}
\caption{\label{fig:cum_mass}
Schematic diagram showing the cumulative stellar mass growth over time (bottom panel) based on the corresponding star formation histories (top panel) typical for galaxies with final masses shown on the right axis of the bottom panel.  High-mass galaxies experience exponentially declining star formation histories with early formation redshifts and short formation timescales.  Low-mass galaxies experience bursty stellar growth behaviour.   }
\end{figure}
%-----   --------------------- FIGURE-----------------------------%

It is likely that low-mass galaxies undergo more bursty episodes of star formation, and on short time scales \citep[e.g.][]{Mateo1998,Dolphin2000,Tolstoy2009,Weisz2011}.  In the intermediate mass range, such as that probed by our volume-limited redshift sample, exponentially declining SFHs do not work well, and a few simple tests show that pure bursty behaviour does not produce the high SSFR of observed low-mass galaxies either.   

For example, take a low-mass galaxy at $z=0.32$ forming stars at a moderate SFR  $= 1~\Msun$ yr$^{-1}$ (shown as a large magenta circle in the top panel of Figure~\ref{fig:gama_ssfr_manytracks}).  If that galaxy forms stars at that constant rate until present, it would have formed M$_*$ = $2.8 \times 10^{9}$ $\Msun$ of stars by the present day, represented by the circle in the lower right panel of Figure~\ref{fig:gama_ssfr_manytracks}.  It is unlikely, however, that a galaxy could maintain such a level of star formation over that period of time due to internal winds \citep{Sharp2010}, for example, or a lack of sufficient gas for fuel \citep{Fabello2011}.  

If instead, the same galaxy shown in the top-left panel of Figure~\ref{fig:gama_ssfr_manytracks} followed an exponentially declining SFH instead of a constant SFR, and only formed stars 25\% of the time, the galaxy would evolve to the location of the triangle in lower right panel of Figure~\ref{fig:gama_ssfr_manytracks}.  Note the position of the triangle assumes that the current SFR  is $0.5~\Msun$ yr$^{-1}$, which is actually an upper limit, as the galaxy could have any SFR of 0.5~$\Msun$ yr$^{-1}$ or less when observed.  The galaxy maintains low-mass status, even if its observed SSFR does not reach the levels observed for positive outliers above the median SSFRs for low-mass galaxies.  While we cannot determine whether this galaxy represents and can account for the behaviour of a ``typical'' galaxy of M$_* = 10^{9}$ $\Msun$, because we do not detect the non-star-formers, we can say that reproducing the upper envelope of systems would require bursts of star formation, at higher intensity and at a lower duty cycle potentially, than this nominal example. We note that the existence of this bursty population is not affected by our selection effects.
 
Galaxies of low- to intermediate-masses are produced using duty cycles of exponentially declining SFHs, but to reproduce low-mass galaxies with the highest SSFRs observed, a burst higher than dictated by the exponential SFH must be achieved.   The presence of so many of these positive outliers represents evidence for short, event driven star formation behaviours.  Galaxies with even lower masses, dwarf galaxies with stellar masses below 10$^{7}$ to 10$^{8}$ $\Msun$, experience bursts of intense star formation \citep{Lee2009,Weisz2011,Nichols2012}, with little evidence of an underlying level of continuous star formation. The bulk of old stars in dwarf galaxies were formed at $z>1$ and then additional stellar populations, adding up to 15\% of the total stellar mass, have been formed any time over the last 1 billion years \citep{Weisz2011}.  

%-----  --------------------- FIGURE-----------------------------%
\begin{figure}
\includegraphics[scale=0.6]{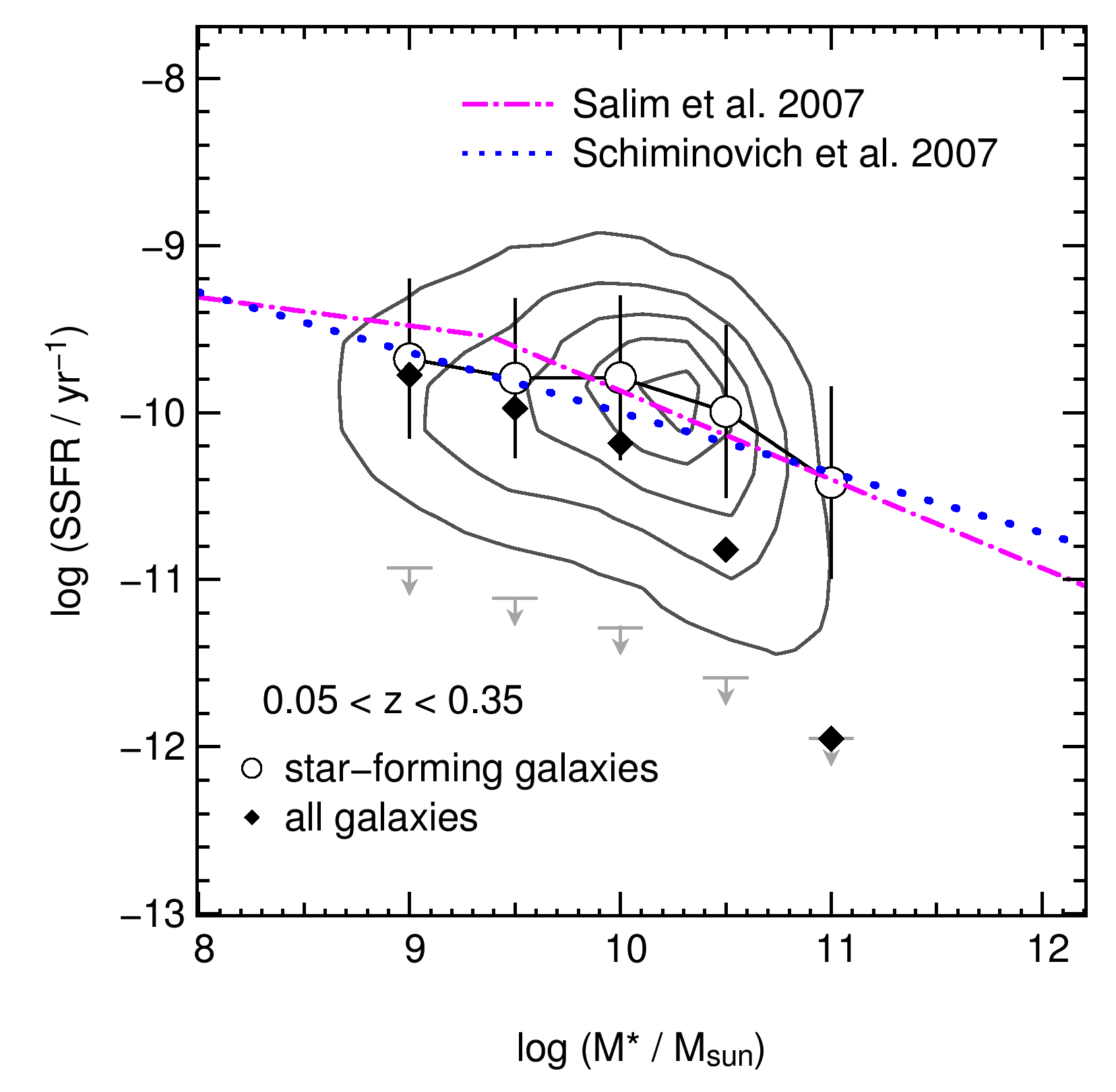}
\caption{\label{fig:gama_ssfr_salim}
SSFR as a function of stellar mass for the full redshift range, similar to Figure~\ref{fig:ssfr_mass_zcolors}.  The magenta dash-dot line and blue dotted lines are the fits of \citet{Salim2007} and \citet{Schim2007}, respectively, to star-forming SDSS galaxies.    }
\end{figure}
%-----   --------------------- FIGURE-----------------------------%

These mass dependent star formation histories and the resulting  cumulative growth of stellar mass are presented in schematic form in Figure~\ref{fig:cum_mass}.  Dwarf galaxies are shown as undergoing burst-driven stellar mass growth, producing the step function of cumulative growth shown in the bottom panel (gray) in Figure~\ref{fig:cum_mass}.  Low-mass galaxies achieve a late onset of exponentially declining SFRs with small bursts superimposed.  In this scenario, it is advantageous that the redshift bins in Figures~\ref{fig:gama_ssfr}~and~\ref{fig:gama_ssfr_manytracks} are not complete in stellar mass because the mass completeness limit is at a high stellar mass (for example, the mass limit for the $0.17<z<0.2$ bin is M$_* = 3\times10^{10}$ $\Msun$), therefore excluding the low-mass galaxies which would not be detectable when not experiencing elevated star formation.  More massive galaxies, on average, can be adequately explained by exponentially declining SFHs with relatively short timescales \citep{Drory2008,Wuyts2011}.  

A possible explanation for the stochastic nature of star formation in low-mass galaxies is the number and distribution of star forming regions within individual galaxies.  The Galaxy, for example, has over 500 giant molecular clouds (GMC) \citep{Roman-Duval2010}.  While these GMCs are not all forming stars at the same time, the sum of all the star formation from the ensemble of clouds would appear to an external observer as a steady rate, not stochastic.  Furthermore, the summation of star formation over these regions would gradually decrease as internal pressure decreases and more individual regions slowly end star formation over time \citep{McKee2007,KennicuttEvans2012}.  In fact recent models have shown that the physics of star formation and stellar feedback are better determined from regions of dense gas in GMCs, as opposed to global galaxy properties such as the Toomre Q parameter or gas velocity dispersion \citep{PhilHopkins2012,Calzetti2012}.  This would account for the slow decline of overall star formation measured in high-mass galaxies at any redshift since the measured SFR is a global value.  

On the other hand, low-mass galaxies have fewer individual star-forming regions and higher neutral gas fractions \citep{LSanchez2010,Fabello2011,LaraLopez2013}.   Therefore, the overall star formation of a low-mass galaxy appears more stochastic in nature, as individual molecular clouds shine and fade during bouts of star formation \citep{Kroupa2011}, as opposed to having a dominating component of widespread, regulated star formation common to more massive systems.

\subsection{Comparison to SDSS}\label{sec:sfh_compare}

The results of the SSFR as a function of stellar mass for the GAMA survey are supported by the work of \citet{Salim2007} and \citet{Schim2007}, which presented studies using the SDSS.   Figure~\ref{fig:gama_ssfr_salim} is similar to Figure~\ref{fig:ssfr_mass_zcolors}, with the addition of relations derived from SDSS for star-forming and composite AGN + star-forming sources \citep{Salim2007}, and for SDSS galaxies using the UV as a star formation rate indicator \citep{Schim2007}.   These studies both find that SSFR is a function of stellar mass for  star-forming galaxies and their derived relation is in excellent agreement with the median values we calculate for GAMA star-forming galaxies.  It can clearly be seen, though, that the behaviour of these models overestimates the SSFR for the highest mass systems.  

A possible alternative solution for SFHs of galaxies is presented in the work of \citet{peng2010}.  \citet{peng2010} based their models on the assumption that the relationship between SSFR and stellar mass is flat over time, citing the relations for SDSS galaxies presented by \citet{Salim2007} and \citet{Elbaz2007}.   As can be seen in Figure~\ref{fig:gama_ssfr_salim}, our data do not support this assumption, and indeed even the \citet{Salim2007} and \citet{Schim2007} observations decline by an order of magnitude in SSFR over the observed range of stellar mass.  For this reason, we do not compare the SFH models derived by \citet{peng2010} with the models presented here.  Overall, we find that SSFR is a strong function of stellar mass up to $z=0.32$ for the population of star-forming galaxies, and an even stronger function still when the full galaxy population is accounted for.

%%%%%%%%%%%%%%%%%%%%%%%%%%%%%%%%%%%%%%%%%%%%                   
Summary and Conclusions         %%
%%%%%%%%%%%%%%%%%%%%%%%%%%%%%%%%%%%%%%%%%%
\section{Conclusions}\label{sec:conclusion}

We present star formation rates (SFR) and specific star formation rates (SSFR) as a function of stellar mass and redshift for $\sim$73,000 GAMA galaxies between $0.05<z<0.32$.  We calculate SFRs from spectroscopic H$\alpha$ measurements and derive dust corrections from Balmer decrements.  We find that SFRs increase as a function of stellar mass up to $3\times10^{10}$ $\Msun$, flatten and then decrease, independent of the dust correction, which depends more strongly on the SFR than the stellar mass.  

We find that SSFR decreases as a function of increasing stellar mass at all redshifts up to $z=0.32$ --- i.e. this relation is not flat.  The dependence of SSFR on stellar mass  is stronger for the full sample of galaxies than when calculated using just star-forming galaxies.   We find that $\sim$70\% of $M_{*}<10^{10}$ $\Msun$ galaxies are forming stars.  This fraction steadily decreases with increasing stellar mass, as an increasing fraction of high mass galaxies show little or no evidence of star formation via H$\alpha$ emission.   We find that for galaxies with $M_{*}>10^{11}$ $\Msun$, only 20\% are forming stars.  From the full sample, 3\% of the galaxies show characteristics of AGN or LINER activity.   We also find that the median SSFR for $M_{*}>10^{10.5}$ $\Msun$ galaxies decreases by a factor of 4 between $z=0.3$ and $z=0.1$.  

Low-mass galaxies exhibit high SSFRs at all redshifts up to $z=0.32$. We use simple parametrizations of star formation histories to investigate the dependence on stellar mass of the star formation timescale (i.e. the $e$-folding time, $\tau$) and the formation redshift.  We find that observed GAMA galaxies have higher SSFRs than predicted by simple models derived from $z\sim1$ galaxies.  The best mass-dependent parametrizations of exponentially declining star formation histories for GAMA cannot sufficiently recover the dependence of SSFRs on stellar mass over the full mass and redshift range of $0.05<z<0.32$.  In particular, we find a population of low mass galaxies that exhibit higher SSFRs than can be achieved by the models, and form too much stellar mass under the prescription of exponentially declining SFHs.   

Reproduction of these low-mass galaxies with the highest SSFRs observed requires the late onset of an underlying exponentially declining SFH with stochastic bursts of star formation superimposed.  The observed galaxies of intermediate-mass can be produced using duty cycles of 25 to 50\% for exponentially declining SFHs with late onset of star formation.    The observed populations of massive GAMA galaxies, on average,  consistent with exponentially declining SFHs with relatively short timescales.  

This behaviour can be accounted for by the presence of individual star-forming regions inside galaxies that combine to produce a steady exponential star formation history for high mass galaxies that harbour hundreds of star-forming regions.  The underlying star formation in individual star-forming regions throughout galaxies is a stochastic process, which becomes evident in low-mass galaxies with fewer individual star formation regions.

\section*{Acknowledgments}
AEB acknowledges the Australian Research Council (ARC) and Super Science Fellowship funding for supporting this work [FS100100065].  JL acknowledges support from the Science and Technology Facilities Council [grant number ST/I000976/1].  CF acknowledges co-funding under the Marie Curie Actions of the European Commission (FP7-COFUND).  

GAMA is a joint European-Australasian project based around a spectroscopic campaign using the Anglo-Australian Telescope. The GAMA input catalogue is based on data taken from the Sloan Digital Sky Survey and the UKIRT Infrared Deep Sky Survey. Complementary imaging of the GAMA regions is being obtained by a number of independent survey programs including GALEX MIS, VST KIDS, VISTA VIKING, WISE, Herschel-ATLAS, GMRT and ASKAP providing UV to radio coverage. GAMA is funded by the STFC (UK), the ARC (Australia), the AAO, and the participating institutions. The GAMA website is http://www.gama-survey.org/. 

Funding for the SDSS and SDSS-II has been provided by the Alfred P. Sloan Foundation, the Participating Institutions, the National Science Foundation, the U.S. Department of Energy, the National Aeronautics and Space Administration, the Japanese Monbukagakusho, the Max Planck Society, and the Higher Education Funding Council for England. The SDSS Web Site is http://www.sdss.org/.

The SDSS is managed by the Astrophysical Research Consortium for the Participating Institutions. The Participating Institutions are the American Museum of Natural History, Astrophysical Institute Potsdam, University of Basel, University of Cambridge, Case Western Reserve University, University of Chicago, Drexel University, Fermilab, the Institute for Advanced Study, the Japan Participation Group, Johns Hopkins University, the Joint Institute for Nuclear Astrophysics, the Kavli Institute for Particle Astrophysics and Cosmology, the Korean Scientist Group, the Chinese Academy of Sciences (LAMOST), Los Alamos National Laboratory, the Max-Planck-Institute for Astronomy (MPIA), the Max-Planck-Institute for Astrophysics (MPA), New Mexico State University, Ohio State University, University of Pittsburgh, University of Portsmouth, Princeton University, the United States Naval Observatory, and the University of Washington.

\bibliographystyle{mn2e}
\bibliography{gama} 

\begin{thebibliography}{91}
\expandafter\ifx\csname natexlab\endcsname\relax\def\natexlab#1{#1}\fi

\bibitem[{{Abazajian} {et~al}\mbox{.}(2009){Abazajian}, {Adelman-McCarthy},
  {Ag{\"u}eros}, {Allam}, {Allende Prieto}, {An}, {Anderson}, {Anderson},
  {Annis}, {Bahcall}, \& et~al.}]{DR72009}
{Abazajian} K.~N. {et~al.}, 2009, ApJS, 182, 543

\bibitem[{{Baldry} {et~al}\mbox{.}(2012){Baldry}, {Driver}, {Loveday},
  {Taylor}, {Kelvin}, {Liske}, {Norberg}, {Robotham}, {Brough}, {Hopkins},
  {Bamford}, {Peacock}, {Bland-Hawthorn}, {Conselice}, {Croom}, {Jones},
  {Parkinson}, {Popescu}, {Prescott}, {Sharp}, \& {Tuffs}}]{Baldry2012}
{Baldry} I.~K. {et~al.}, 2012, MNRAS, 421, 621

\bibitem[{{Baldry} {et~al}\mbox{.}(2008){Baldry}, {Glazebrook}, \&
  {Driver}}]{Baldry2008}
{Baldry} I.~K., {Glazebrook} K., {Driver} S.~P., 2008, MNRAS, 388, 945

\bibitem[{{Baldry} {et~al}\mbox{.}(2010){Baldry}, {Robotham}, {Hill}, {Driver},
  {Liske}, {Norberg}, {Bamford}, {Hopkins}, {Loveday}, {Peacock}, {Cameron},
  {Croom}, {Cross}, {Doyle}, {Dye}, {Frenk}, {Jones}, \& {van
  Kampen}}]{Baldry2010}
{Baldry} I.~K. {et~al.}, 2010, MNRAS, 404, 86

\bibitem[{{Baldwin} {et~al}\mbox{.}(1981){Baldwin}, {Phillips}, \&
  {Terlevich}}]{BPT1981}
{Baldwin} J.~A., {Phillips} M.~M., {Terlevich} R., 1981, PASP, 93, 5

\bibitem[{{Bauer} {et~al}\mbox{.}(2011){Bauer}, {Conselice},
  {P{\'e}rez-Gonz{\'a}lez}, {Gr{\"u}tzbauch}, {Bluck}, {Buitrago}, \&
  {Mortlock}}]{Bauer2011_GNS}
{Bauer} A.~E., {Conselice} C.~J., {P{\'e}rez-Gonz{\'a}lez} P.~G.,
  {Gr{\"u}tzbauch} R., {Bluck} A.~F.~L., {Buitrago} F., {Mortlock} A., 2011,
  MNRAS, 417, 289

\bibitem[{{Bauer} {et~al}\mbox{.}(2005){Bauer}, {Drory}, {Hill}, \&
  {Feulner}}]{Bauer05}
{Bauer} A.~E., {Drory} N., {Hill} G.~J., {Feulner} G., 2005, ApJL, 621, L89

\bibitem[{{Bell} {et~al}\mbox{.}(2003){Bell}, {McIntosh}, {Katz}, \&
  {Weinberg}}]{Bell2003}
{Bell} E.~F., {McIntosh} D.~H., {Katz} N., {Weinberg} M.~D., 2003, ApJS, 149,
  289

\bibitem[{{Bell} {et~al}\mbox{.}(2007){Bell}, {Zheng}, {Papovich}, {Borch},
  {Wolf}, \& {Meisenheimer}}]{Bell2007}
{Bell} E.~F., {Zheng} X.~Z., {Papovich} C., {Borch} A., {Wolf} C.,
  {Meisenheimer} K., 2007, ApJ, 663, 834

\bibitem[{{Bolzonella} {et~al}\mbox{.}(2010){Bolzonella}, {Kova{\v c}},
  {Pozzetti}, {Zucca}, {Cucciati}, {Lilly}, {Peng}, {Iovino}, {Zamorani},
  {Vergani}, {Tasca}, {Lamareille}, {Oesch}, {Caputi}, {Kampczyk}, {Bardelli},
  {Maier}, {Abbas}, {Knobel}, {Scodeggio}, {Carollo}, {Contini}, {Kneib}, {Le
  F{\`e}vre}, {Mainieri}, {Renzini}, {Bongiorno}, {Coppa}, {de la Torre}, {de
  Ravel}, {Franzetti}, {Garilli}, {Le Borgne}, {Le Brun}, {Mignoli},
  {Pell{\'o}}, {Perez-Montero}, {Ricciardelli}, {Silverman}, {Tanaka},
  {Tresse}, {Bottini}, {Cappi}, {Cassata}, {Cimatti}, {Guzzo}, {Koekemoer},
  {Leauthaud}, {Maccagni}, {Marinoni}, {McCracken}, {Memeo}, {Meneux},
  {Porciani}, {Scaramella}, {Aussel}, {Capak}, {Halliday}, {Ilbert},
  {Kartaltepe}, {Salvato}, {Sanders}, {Scarlata}, {Scoville}, {Taniguchi}, \&
  {Thompson}}]{Bolzonella2010}
{Bolzonella} M. {et~al.}, 2010, A\&A, 524, A76

\bibitem[{{Bower} {et~al}\mbox{.}(2006){Bower}, {Benson}, {Malbon}, {Helly},
  {Frenk}, {Baugh}, {Cole}, \& {Lacey}}]{Bower2006}
{Bower} R.~G., {Benson} A.~J., {Malbon} R., {Helly} J.~C., {Frenk} C.~S.,
  {Baugh} C.~M., {Cole} S., {Lacey} C.~G., 2006, MNRAS, 370, 645

\bibitem[{{Brough} {et~al}\mbox{.}(2011){Brough}, {Hopkins}, {Sharp},
  {Gunawardhana}, {Wijesinghe}, {Robotham}, {Driver}, {Baldry}, {Bamford},
  {Liske}, {Loveday}, {Norberg}, {Peacock}, {Bland-Hawthorn}, {Brown},
  {Cameron}, {Croom}, {Frenk}, {Foster}, {Hill}, {Jones}, {Kelvin}, {Kuijken},
  {Nichol}, {Parkinson}, {Pimbblet}, {Popescu}, {Prescott}, {Sutherland},
  {Taylor}, {Thomas}, {Tuffs}, \& {van Kampen}}]{Brough2011}
{Brough} S. {et~al.}, 2011, MNRAS, 413, 1236

\bibitem[{{Bruzual} \& {Charlot}(2003)}]{BC03}
{Bruzual} G., {Charlot} S., 2003, MNRAS, 344, 1000

\bibitem[{{Bundy} {et~al}\mbox{.}(2006){Bundy}, {Ellis}, {Conselice}, {Taylor},
  {Cooper}, {Willmer}, {Weiner}, {Coil}, {Noeske}, \& {Eisenhardt}}]{Bundy2006}
{Bundy} K. {et~al.}, 2006, ApJ, 651, 120

\bibitem[{{Calzetti} {et~al}\mbox{.}(2000){Calzetti}, {Armus}, {Bohlin},
  {Kinney}, {Koornneef}, \& {Storchi-Bergmann}}]{Calzetti2000}
{Calzetti} D., {Armus} L., {Bohlin} R.~C., {Kinney} A.~L., {Koornneef} J.,
  {Storchi-Bergmann} T., 2000, ApJ, 533, 682

\bibitem[{{Calzetti} {et~al}\mbox{.}(2012){Calzetti}, {Liu}, \&
  {Koda}}]{Calzetti2012}
{Calzetti} D., {Liu} G., {Koda} J., 2012, ArXiv e-prints

\bibitem[{{Caputi} {et~al}\mbox{.}(2008){Caputi}, {Lilly}, {Aussel}, {Sanders},
  {Frayer}, {Le F{\`e}vre}, {Renzini}, {Zamorani}, {Scodeggio}, {Contini},
  {Scoville}, {Carollo}, {Hasinger}, {Iovino}, {Le Brun}, {Le Floc'h}, {Maier},
  {Mainieri}, {Mignoli}, {Salvato}, {Schiminovich}, {Silverman}, {Surace},
  {Tasca}, {Abbas}, {Bardelli}, {Bolzonella}, {Bongiorno}, {Bottini}, {Capak},
  {Cappi}, {Cassata}, {Cimatti}, {Cucciati}, {de la Torre}, {de Ravel},
  {Franzetti}, {Fumana}, {Garilli}, {Halliday}, {Ilbert}, {Kampczyk},
  {Kartaltepe}, {Kneib}, {Knobel}, {Kovac}, {Lamareille}, {Leauthaud}, {Le
  Borgne}, {Maccagni}, {Marinoni}, {McCracken}, {Meneux}, {Oesch}, {Pell{\`o}},
  {P{\'e}rez-Montero}, {Porciani}, {Ricciardelli}, {Scaramella}, {Scarlata},
  {Tresse}, {Vergani}, {Walcher}, {Zamojski}, \& {Zucca}}]{Caputi2008}
{Caputi} K.~I. {et~al.}, 2008, ApJ, 680, 939

\bibitem[{{Cardamone} {et~al}\mbox{.}(2009){Cardamone}, {Schawinski}, {Sarzi},
  {Bamford}, {Bennert}, {Urry}, {Lintott}, {Keel}, {Parejko}, {Nichol},
  {Thomas}, {Andreescu}, {Murray}, {Raddick}, {Slosar}, {Szalay}, \&
  {Vandenberg}}]{Peas}
{Cardamone} C. {et~al.}, 2009, MNRAS, 399, 1191

\bibitem[{{Chabrier}(2003)}]{Chabrier2003}
{Chabrier} G., 2003, PASP, 115, 763

\bibitem[{{Cole} {et~al}\mbox{.}(2001){Cole}, {Norberg}, {Baugh}, {Frenk},
  {Bland-Hawthorn}, {Bridges}, {Cannon}, {Colless}, {Collins}, {Couch},
  {Cross}, {Dalton}, \& {De Propris}}]{Cole2001}
{Cole} S. {et~al.}, 2001, MNRAS, 326, 255

\bibitem[{{Cowie} {et~al}\mbox{.}(1996){Cowie}, {Songaila}, {Hu}, \&
  {Cohen}}]{Cowie1996}
{Cowie} L.~L., {Songaila} A., {Hu} E.~M., {Cohen} J.~G., 1996, AJ, 112, 839

\bibitem[{{Croton} {et~al}\mbox{.}(2006){Croton}, {Springel}, {White}, {De
  Lucia}, {Frenk}, {Gao}, {Jenkins}, {Kauffmann}, {Navarro}, \&
  {Yoshida}}]{Croton2006}
{Croton} D.~J. {et~al.}, 2006, MNRAS, 365, 11

\bibitem[{{Daddi} {et~al}\mbox{.}(2007){Daddi}, {Dickinson}, {Morrison},
  {Chary}, {Cimatti}, {Elbaz}, {Frayer}, {Renzini}, {Pope}, {Alexander},
  {Bauer}, {Giavalisco}, {Huynh}, {Kurk}, \& {Mignoli}}]{Daddi2007}
{Daddi} E. {et~al.}, 2007, ApJ, 670, 156

\bibitem[{{Damen} {et~al}\mbox{.}(2009){Damen}, {Labb{\'e}}, {Franx}, {van
  Dokkum}, {Taylor}, \& {Gawiser}}]{Damen2009}
{Damen} M., {Labb{\'e}} I., {Franx} M., {van Dokkum} P.~G., {Taylor} E.~N.,
  {Gawiser} E.~J., 2009, ApJ, 690, 937

\bibitem[{{Dav{\'e}}(2008)}]{Dave2008}
{Dav{\'e}} R., 2008, MNRAS, 385, 147

\bibitem[{{De Lucia} \& {Blaizot}(2007)}]{DLB2007}
{De Lucia} G., {Blaizot} J., 2007, MNRAS, 375, 2

\bibitem[{{Dickinson} {et~al}\mbox{.}(2003){Dickinson}, {Papovich}, {Ferguson},
  \& {Budav{\'a}ri}}]{Dickinson2003}
{Dickinson} M., {Papovich} C., {Ferguson} H.~C., {Budav{\'a}ri} T., 2003, ApJ,
  587, 25

\bibitem[{{Dolphin}(2000)}]{Dolphin2000}
{Dolphin} A.~E., 2000, MNRAS, 313, 281

\bibitem[{{Driver} {et~al}\mbox{.}(2011){Driver}, {Hill}, {Kelvin}, {Robotham},
  {Liske}, {Norberg}, {Baldry}, {Bamford}, {Hopkins}, {Loveday}, {Peacock},
  {Andrae}, {Bland-Hawthorn}, {Brough}, {Brown}, {Cameron}, {Ching}, \&
  {Colless}}]{Driver2011}
{Driver} S.~P. {et~al.}, 2011, MNRAS, 413, 971

\bibitem[{{Drory} \& {Alvarez}(2008)}]{Drory2008}
{Drory} N., {Alvarez} M., 2008, ApJ, 680, 41

\bibitem[{{Drory} {et~al}\mbox{.}(2009){Drory}, {Bundy}, {Leauthaud},
  {Scoville}, {Capak}, {Ilbert}, {Kartaltepe}, {Kneib}, {McCracken}, {Salvato},
  {Sanders}, {Thompson}, \& {Willott}}]{Drory2009}
{Drory} N. {et~al.}, 2009, ApJ, 707, 1595

\bibitem[{{Drory} {et~al}\mbox{.}(2005){Drory}, {Salvato}, {Gabasch}, {Bender},
  {Hopp}, {Feulner}, \& {Pannella}}]{Drory2005}
{Drory} N., {Salvato} M., {Gabasch} A., {Bender} R., {Hopp} U., {Feulner} G.,
  {Pannella} M., 2005, ApJL, 619, L131

\bibitem[{{Elbaz} {et~al}\mbox{.}(2007){Elbaz}, {Daddi}, {Le Borgne},
  {Dickinson}, {Alexander}, {Chary}, {Starck}, {Brandt}, {Kitzbichler},
  {MacDonald}, {Nonino}, {Popesso}, {Stern}, \& {Vanzella}}]{Elbaz2007}
{Elbaz} D. {et~al.}, 2007, AA, 468, 33

\bibitem[{{Fabello} {et~al}\mbox{.}(2011){Fabello}, {Catinella}, {Giovanelli},
  {Kauffmann}, {Haynes}, {Heckman}, \& {Schiminovich}}]{Fabello2011}
{Fabello} S., {Catinella} B., {Giovanelli} R., {Kauffmann} G., {Haynes} M.~P.,
  {Heckman} T.~M., {Schiminovich} D., 2011, MNRAS, 411, 993

\bibitem[{{Fontana} {et~al}\mbox{.}(2003){Fontana}, {Donnarumma}, {Vanzella},
  {Giallongo}, {Menci}, {Nonino}, {Saracco}, {Cristiani}, {D'Odorico}, \&
  {Poli}}]{Fontana2003}
{Fontana} A. {et~al.}, 2003, ApJL, 594, L9

\bibitem[{{Fontanot} {et~al}\mbox{.}(2012){Fontanot}, {Cristiani}, {Santini},
  {Fontana}, {Grazian}, \& {Somerville}}]{Fontanot2012}
{Fontanot} F., {Cristiani} S., {Santini} P., {Fontana} A., {Grazian} A.,
  {Somerville} R.~S., 2012, MNRAS, 421, 241

\bibitem[{{Gilbank} {et~al}\mbox{.}(2011){Gilbank}, {Bower}, {Glazebrook},
  {Balogh}, {Baldry}, {Davies}, {Hau}, {Li}, {McCarthy}, \&
  {Sawicki}}]{Gilbank2011}
{Gilbank} D.~G. {et~al.}, 2011, MNRAS, 414, 304

\bibitem[{{Governato} {et~al}\mbox{.}(2007){Governato}, {Willman}, {Mayer},
  {Brooks}, {Stinson}, {Valenzuela}, {Wadsley}, \& {Quinn}}]{Governato2007}
{Governato} F., {Willman} B., {Mayer} L., {Brooks} A., {Stinson} G.,
  {Valenzuela} O., {Wadsley} J., {Quinn} T., 2007, MNRAS, 374, 1479

\bibitem[{{Gunawardhana} {et~al}\mbox{.}(2011){Gunawardhana}, {Hopkins},
  {Sharp}, {Brough}, {Taylor}, {Bland-Hawthorn}, {Maraston}, {Tuffs},
  {Popescu}, {Wijesinghe}, {Jones}, {Croom}, {Sadler}, {Wilkins}, {Driver},
  {Liske}, \& {Norberg}}]{Madusha2011}
{Gunawardhana} M.~L.~P. {et~al.}, 2011, MNRAS, 415, 1647

\bibitem[{{Gunawardhana} {et~al}\mbox{.}(2013){Gunawardhana}, {Hopkins},
  {Sharp}, {Brough}, {Taylor}, {Bland-Hawthorn}, {Maraston}, {Tuffs},
  {Popescu}, {Wijesinghe}, {Jones}, {Croom}, {Sadler}, {Wilkins}, {Driver},
  {Liske}, \& {Norberg}}]{Madusha2013}
{Gunawardhana} M.~L.~P. {et~al.}, 2013, MNRAS, submitted

\bibitem[{{Hill} {et~al}\mbox{.}(2011){Hill}, {Kelvin}, {Driver}, {Robotham},
  {Cameron}, {Cross}, {Andrae}, {Baldry}, {Bamford}, {Bland-Hawthorn},
  {Brough}, {Conselice}, {Dye}, {Hopkins}, {Liske}, {Loveday}, {Norberg},
  {Peacock}, {Croom}, {Frenk}, {Graham}, {Jones}, {Kuijken}, {Madore},
  {Nichol}, {Parkinson}, {Phillipps}, {Pimbblet}, {Popescu}, {Prescott},
  {Seibert}, {Sharp}, {Sutherland}, {Thomas}, {Tuffs}, \& {van
  Kampen}}]{Hill2011}
{Hill} D.~T. {et~al.}, 2011, MNRAS, 412, 765

\bibitem[{{Hopkins} \& {Beacom}(2006)}]{HB2006}
{Hopkins} A.~M., {Beacom} J.~F., 2006, ApJ, 651, 142

\bibitem[{{Hopkins} {et~al}\mbox{.}(2013){Hopkins}, {Driver}, {Brough},
  {Owers}, {Bauer}, {Gunawardhana}, {Cluver}, {Colless}, {Foster},
  {Lara-L{\'o}pez}, {Roseboom}, {Sharp}, {Steele}, {Thomas}, {Baldry}, {Brown},
  {Liske}, {Norberg}, {Robotham}, {Bamford}, {Bland-Hawthorn}, {Drinkwater},
  {Loveday}, {Meyer}, {Peacock}, {Tuffs}, {Agius}, {Alpaslan}, {Andrae},
  {Cameron}, {Cole}, {Ching}, {Christodoulou}, {Conselice}, {Croom}, {Cross},
  {De Propris}, {Delhaize}, {Dunne}, {Eales}, {Ellis}, {Frenk}, {Graham},
  {Grootes}, {H{\"a}u{\ss}ler}, {Heymans}, {Hill}, {Hoyle}, {Hudson}, {Jarvis},
  {Johansson}, {Jones}, {van Kampen}, {Kelvin}, {Kuijken},
  {L{\'o}pez-S{\'a}nchez}, {Maddox}, {Madore}, {Maraston}, {McNaught-Roberts},
  {Nichol}, {Oliver}, {Parkinson}, {Penny}, {Phillipps}, {Pimbblet}, {Ponman},
  {Popescu}, {Prescott}, {Proctor}, {Sadler}, {Sansom}, {Seibert},
  {Staveley-Smith}, {Sutherland}, {Taylor}, {Van Waerbeke}, {V{\'a}zquez-Mata},
  {Warren}, {Wijesinghe}, {Wild}, \& {Wilkins}}]{hopkins2013}
{Hopkins} A.~M. {et~al.}, 2013, MNRAS, 430, 2047

\bibitem[{{Hopkins} {et~al}\mbox{.}(2003){Hopkins}, {Miller}, {Nichol},
  {Connolly}, {Bernardi}, {G{\'o}mez}, {Goto}, {Tremonti}, {Brinkmann},
  {Ivezi{\'c}}, \& {Lamb}}]{Hopkins2003}
{Hopkins} A.~M. {et~al.}, 2003, ApJ, 599, 971

\bibitem[{{Hopkins} {et~al}\mbox{.}(2012){Hopkins}, {Quataert}, \&
  {Murray}}]{PhilHopkins2012}
{Hopkins} P.~F., {Quataert} E., {Murray} N., 2012, MNRAS, 421, 3488

\bibitem[{{Ilbert} {et~al}\mbox{.}(2010){Ilbert}, {Salvato}, {Le Floc'h},
  {Aussel}, {Capak}, {McCracken}, {Mobasher}, {Kartaltepe}, {Scoville},
  {Sanders}, {Arnouts}, {Bundy}, {Cassata}, {Kneib}, {Koekemoer}, {Le
  F{\`e}vre}, {Lilly}, {Surace}, {Taniguchi}, {Tasca}, {Thompson}, {Tresse},
  {Zamojski}, {Zamorani}, \& {Zucca}}]{Ilbert2010}
{Ilbert} O. {et~al.}, 2010, ApJ, 709, 644

\bibitem[{{Karim} {et~al}\mbox{.}(2011){Karim}, {Schinnerer},
  {Mart{\'{\i}}nez-Sansigre}, {Sargent}, {van der Wel}, {Rix}, {Ilbert},
  {Smol{\v c}i{\'c}}, {Carilli}, {Pannella}, {Koekemoer}, {Bell}, \&
  {Salvato}}]{Karim2011}
{Karim} A. {et~al.}, 2011, ApJ, 730, 61

\bibitem[{{Kennicutt}(1998)}]{Kennicutt1998}
{Kennicutt}, Jr. R.~C., 1998, ARAA, 36, 189

\bibitem[{{Kennicutt} \& {Evans}(2012)}]{KennicuttEvans2012}
{Kennicutt}, Jr. R.~C., {Evans}, II N.~J., 2012, ArXiv e-prints

\bibitem[{{Kewley} {et~al}\mbox{.}(2001){Kewley}, {Dopita}, {Sutherland},
  {Heisler}, \& {Trevena}}]{Kewley2001}
{Kewley} L.~J., {Dopita} M.~A., {Sutherland} R.~S., {Heisler} C.~A., {Trevena}
  J., 2001, ApJ, 556, 121

\bibitem[{{Kroupa} {et~al}\mbox{.}(2011){Kroupa}, {Weidner},
  {Pflamm-Altenburg}, {Thies}, {Dabringhausen}, {Marks}, \&
  {Maschberger}}]{Kroupa2011}
{Kroupa} P., {Weidner} C., {Pflamm-Altenburg} J., {Thies} I., {Dabringhausen}
  J., {Marks} M., {Maschberger} T., 2011, ArXiv e-prints

\bibitem[{{Lara-L{\'o}pez} {et~al}\mbox{.}(2013){Lara-L{\'o}pez}, {Hopkins},
  {L{\'o}pez-S{\'a}nchez}, {Brough}, {Bland-Hawthorn}, {Driver}, {Foster},
  {Liske}, {Loveday}, {Robotham}, {Sharp}, {Steele}, \&
  {Taylor}}]{LaraLopez2013}
{Lara-L{\'o}pez} M.~A. {et~al.}, 2013, ArXiv e-prints

\bibitem[{{Lee} {et~al}\mbox{.}(2009){Lee}, {Gil de Paz}, {Tremonti},
  {Kennicutt}, {Salim}, {Bothwell}, {Calzetti}, {Dalcanton}, {Dale},
  {Engelbracht}, {Funes}, {Johnson}, {Sakai}, {Skillman}, {van Zee}, {Walter},
  \& {Weisz}}]{Lee2009}
{Lee} J.~C. {et~al.}, 2009, ApJ, 706, 599

\bibitem[{{Leitner}(2012)}]{Leitner2012}
{Leitner} S.~N., 2012, ApJ, 745, 149

\bibitem[{{Li} \& {White}(2009)}]{Li2009}
{Li} C., {White} S.~D.~M., 2009, MNRAS, 398, 2177

\bibitem[{{Lilly} {et~al}\mbox{.}(1996){Lilly}, {Le Fevre}, {Hammer}, \&
  {Crampton}}]{Lilly1996}
{Lilly} S.~J., {Le Fevre} O., {Hammer} F., {Crampton} D., 1996, ApJL, 460, L1+

\bibitem[{{L{\'o}pez-S{\'a}nchez}(2010)}]{LSanchez2010}
{L{\'o}pez-S{\'a}nchez} {\'A}.~R., 2010, A\&A, 512, 63

\bibitem[{{L{\'o}pez-S{\'a}nchez} \& {Esteban}(2009)}]{LSanchez2009}
{L{\'o}pez-S{\'a}nchez} A.~R., {Esteban} C., 2009, A\&A, 508, 615

\bibitem[{{Madau} {et~al}\mbox{.}(1996){Madau}, {Ferguson}, {Dickinson},
  {Giavalisco}, {Steidel}, \& {Fruchter}}]{Madau1996}
{Madau} P., {Ferguson} H.~C., {Dickinson} M.~E., {Giavalisco} M., {Steidel}
  C.~C., {Fruchter} A., 1996, MNRAS, 283, 1388

\bibitem[{{Marchesini} {et~al}\mbox{.}(2012){Marchesini}, {Stefanon},
  {Brammer}, \& {Whitaker}}]{Marchesini2012}
{Marchesini} D., {Stefanon} M., {Brammer} G.~B., {Whitaker} K.~E., 2012, ApJ,
  748, 126

\bibitem[{{Martin} {et~al}\mbox{.}(2007){Martin}, {Small}, {Schiminovich},
  {Wyder}, {P{\'e}rez-Gonz{\'a}lez}, {Johnson}, {Wolf}, {Barlow}, {Forster},
  {Friedman}, {Morrissey}, \& {Neff}}]{Martin07}
{Martin} D.~C. {et~al.}, 2007, ApJS, 173, 415

\bibitem[{{Mateo}(1998)}]{Mateo1998}
{Mateo} M.~L., 1998, ARAA, 36, 435

\bibitem[{{McKee} \& {Ostriker}(2007)}]{McKee2007}
{McKee} C.~F., {Ostriker} E.~C., 2007, ARAA, 45, 565

\bibitem[{{Nichols} {et~al}\mbox{.}(2012){Nichols}, {Lin}, \&
  {Bland-Hawthorn}}]{Nichols2012}
{Nichols} M., {Lin} D., {Bland-Hawthorn} J., 2012, ApJ, 748, 149

\bibitem[{{Noeske}(2009)}]{Noeske2009}
{Noeske} K.~G., 2009, in Astronomical Society of the Pacific Conference Series,
  Vol. 419, Galaxy Evolution: Emerging Insights and Future Challenges,
  {S.~Jogee, I.~Marinova, L.~Hao, \& G.~A.~Blanc}, ed., p. 298

\bibitem[{{Noeske} {et~al}\mbox{.}(2007{\natexlab{a}}){Noeske}, {Faber},
  {Weiner}, {Koo}, {Primack}, {Dekel}, {Papovich}, {Conselice}, {Le Floc'h},
  {Rieke}, {Coil}, {Lotz}, {Somerville}, \& {Bundy}}]{Noeske07_tau}
{Noeske} K.~G. {et~al.}, 2007{\natexlab{a}}, ApJL, 660, L47

\bibitem[{{Noeske} {et~al}\mbox{.}(2007{\natexlab{b}}){Noeske}, {Weiner},
  {Faber}, {Papovich}, {Koo}, {Somerville}, {Bundy}, {Conselice}, {Newman},
  {Schiminovich}, {Le Floc'h}, {Coil}, {Rieke}, \& {Lotz}}]{Noeske2007_MS}
{Noeske} K.~G. {et~al.}, 2007{\natexlab{b}}, ApJL, 660, L43

\bibitem[{{Oliver} {et~al}\mbox{.}(2010){Oliver}, {Frost}, {Farrah},
  {Gonzalez-Solares}, {Shupe}, {Henriques}, {Roseboom}, {Alfonso-Luis},
  {Babbedge}, {Frayer}, {Lencz}, {Lonsdale}, {Masci}, {Padgett}, {Polletta},
  {Rowan-Robinson}, {Siana}, {Smith}, {Surace}, \& {Vaccari}}]{Oliver2010}
{Oliver} S. {et~al.}, 2010, MNRAS, 405, 2279

\bibitem[{{Peng} {et~al}\mbox{.}(2010){Peng}, {Lilly}, {Kova{\v c}},
  {Bolzonella}, {Pozzetti}, {Renzini}, {Zamorani}, {Ilbert}, {Knobel},
  {Iovino}, {Maier}, {Cucciati}, {Tasca}, {Carollo}, {Silverman}, {Kampczyk},
  {de Ravel}, {Saners}, {Scoville}, {Contini}, {Mainieri}, {Scodeggio},
  {Kneib}, {Le F{\`e}vre}, {Bardelli}, {Bongiorno}, {Caputi}, {Coppa}, {de la
  Torre}, {Franzetti}, {Garilli}, {Lamareille}, {Le Borgne}, {Le Brun},
  {Mignoli}, {Perez Montero}, {Pello}, {Ricciardelli}, {Tanaka}, {Tresse},
  {Vergani}, {Welikala}, {Zucca}, {Oesch}, {Abbas}, {Barnes}, {Bordoloi},
  {Bottini}, {Cappi}, {Cassata}, {Cimatti}, {Fumana}, {Hasinger}, {Koekemoer},
  {Leauthaud}, {Maccagni}, {Marinoni}, {McCracken}, {Memeo}, {Meneux}, {Nair},
  {Porciani}, {Presotto}, \& {Scaramella}}]{peng2010}
{Peng} Y.-j. {et~al.}, 2010, ApJ, 721, 193

\bibitem[{{P{\'e}rez-Gonz{\'a}lez}
  {et~al}\mbox{.}(2008){P{\'e}rez-Gonz{\'a}lez}, {Rieke}, {Villar}, {Barro},
  {Blaylock}, {Egami}, {Gallego}, {Gil de Paz}, {Pascual}, {Zamorano}, \&
  {Donley}}]{PG2008}
{P{\'e}rez-Gonz{\'a}lez} P.~G. {et~al.}, 2008, ApJ, 675, 234

\bibitem[{{Pozzetti} {et~al}\mbox{.}(2010){Pozzetti}, {Bolzonella}, {Zucca},
  {Zamorani}, {Lilly}, {Renzini}, {Moresco}, {Mignoli}, {Cassata}, {Tasca},
  {Lamareille}, {Maier}, {Meneux}, {Halliday}, {Oesch}, {Vergani}, {Caputi},
  {Kova{\v c}}, {Cimatti}, {Cucciati}, {Iovino}, {Peng}, {Carollo}, {Contini},
  {Kneib}, {Le F{\'e}vre}, {Mainieri}, {Scodeggio}, {Bardelli}, {Bongiorno},
  {Coppa}, {de la Torre}, {de Ravel}, {Franzetti}, {Garilli}, {Kampczyk},
  {Knobel}, {Le Borgne}, {Le Brun}, {Pell{\`o}}, {Perez Montero},
  {Ricciardelli}, {Silverman}, {Tanaka}, {Tresse}, {Abbas}, {Bottini}, {Cappi},
  {Guzzo}, {Koekemoer}, {Leauthaud}, {Maccagni}, {Marinoni}, {McCracken},
  {Memeo}, {Porciani}, {Scaramella}, {Scarlata}, \& {Scoville}}]{Pozetti2010}
{Pozzetti} L. {et~al.}, 2010, A\&A, 523, A13

\bibitem[{{Robotham} {et~al}\mbox{.}(2010){Robotham}, {Driver}, {Norberg},
  {Baldry}, {Bamford}, {Hopkins}, {Liske}, {Loveday}, {Peacock}, {Cameron},
  {Croom}, {Doyle}, {Frenk}, \& {Hill}}]{Robotham2010}
{Robotham} A. {et~al.}, 2010, PASA, 27, 76

\bibitem[{{Roman-Duval} {et~al}\mbox{.}(2010){Roman-Duval}, {Jackson}, {Heyer},
  {Rathborne}, \& {Simon}}]{Roman-Duval2010}
{Roman-Duval} J., {Jackson} J.~M., {Heyer} M., {Rathborne} J., {Simon} R.,
  2010, ApJ, 723, 492

\bibitem[{{Rujopakarn} {et~al}\mbox{.}(2010){Rujopakarn}, {Eisenstein},
  {Rieke}, {Papovich}, {Cool}, {Moustakas}, {Jannuzi}, {Kochanek}, {Rieke},
  {Dey}, {Eisenhardt}, {Murray}, {Brown}, \& {Le Floc'h}}]{Rujopakar2010}
{Rujopakarn} W. {et~al.}, 2010, ApJ, 718, 1171

\bibitem[{{Salim} {et~al}\mbox{.}(2007){Salim}, {Rich}, {Charlot},
  {Brinchmann}, {Johnson}, {Schiminovich}, {Seibert}, {Mallery}, {Heckman},
  {Forster}, {Friedman}, {Martin}, \& {Morrissey}}]{Salim2007}
{Salim} S. {et~al.}, 2007, ApJS, 173, 267

\bibitem[{{Santini} {et~al}\mbox{.}(2009){Santini}, {Fontana}, {Grazian},
  {Salimbeni}, {Fiore}, {Fontanot}, {Boutsia}, {Castellano}, {Cristiani}, {de
  Santis}, {Gallozzi}, {Giallongo}, {Menci}, {Nonino}, {Paris}, {Pentericci},
  \& {Vanzella}}]{Santini2009}
{Santini} P. {et~al.}, 2009, AAP, 504, 751

\bibitem[{{Schiminovich} {et~al}\mbox{.}(2007){Schiminovich}, {Wyder},
  {Martin}, {Johnson}, {Salim}, {Seibert}, {Treyer}, {Budav{\'a}ri}, {Hoopes},
  {Zamojski}, {Barlow}, {Forster}, {Friedman}, {Morrissey}, {Neff}, {Small},
  {Bianchi}, {Donas}, {Heckman}, {Lee}, {Madore}, {Milliard}, {Rich}, {Szalay},
  {Welsh}, \& {Yi}}]{Schim2007}
{Schiminovich} D. {et~al.}, 2007, ApJS, 173, 315

\bibitem[{{Sharp} \& {Bland-Hawthorn}(2010)}]{Sharp2010}
{Sharp} R.~G., {Bland-Hawthorn} J., 2010, ApJ, 711, 818

\bibitem[{{Springel} {et~al}\mbox{.}(2006){Springel}, {Frenk}, \&
  {White}}]{Springel2006}
{Springel} V., {Frenk} C.~S., {White} S.~D.~M., 2006, Nature, 440, 1137

\bibitem[{{Taylor} {et~al}\mbox{.}(2011){Taylor}, {Hopkins}, {Baldry}, {Brown},
  {Driver}, {Kelvin}, {Hill}, {Robotham}, {Bland-Hawthorn}, {Jones}, {Sharp},
  {Thomas}, {Liske}, {Loveday}, {Norberg}, {Peacock}, {Bamford}, {Brough},
  {Colless}, {Cameron}, {Conselice}, {Croom}, {Frenk}, {Gunawardhana},
  {Kuijken}, {Nichol}, {Parkinson}, {Phillipps}, {Pimbblet}, {Popescu},
  {Prescott}, {Sutherland}, {Tuffs}, {van Kampen}, \&
  {Wijesinghe}}]{Taylor2011}
{Taylor} E.~N. {et~al.}, 2011, MNRAS, 418, 1587

\bibitem[{{Thomas} {et~al}\mbox{.}(2005){Thomas}, {Maraston}, {Bender}, \&
  {Mendes de Oliveira}}]{Thomas2005}
{Thomas} D., {Maraston} C., {Bender} R., {Mendes de Oliveira} C., 2005, ApJ,
  621, 673

\bibitem[{{Thomas} {et~al}\mbox{.}(2010){Thomas}, {Maraston}, {Schawinski},
  {Sarzi}, \& {Silk}}]{Thomas2011}
{Thomas} D., {Maraston} C., {Schawinski} K., {Sarzi} M., {Silk} J., 2010,
  MNRAS, 404, 1775

\bibitem[{{Tinsley}(1968)}]{Tinsley1968}
{Tinsley} B.~M., 1968, ApJ, 151, 547

\bibitem[{{Tolstoy} {et~al}\mbox{.}(2009){Tolstoy}, {Hill}, \&
  {Tosi}}]{Tolstoy2009}
{Tolstoy} E., {Hill} V., {Tosi} M., 2009, ARAA, 47, 371

\bibitem[{{Weisz} {et~al}\mbox{.}(2011){Weisz}, {Dalcanton}, {Williams},
  {Gilbert}, {Skillman}, {Seth}, {Dolphin}, {McQuinn}, {Gogarten}, {Holtzman},
  {Rosema}, {Cole}, {Karachentsev}, \& {Zaritsky}}]{Weisz2011}
{Weisz} D.~R. {et~al.}, 2011, ApJ, 739, 5

\bibitem[{{Whitaker} {et~al}\mbox{.}(2012){Whitaker}, {van Dokkum}, {Brammer},
  \& {Franx}}]{Whitaker2012}
{Whitaker} K.~E., {van Dokkum} P.~G., {Brammer} G., {Franx} M., 2012, ApJL,
  754, L29

\bibitem[{{Wijesinghe} {et~al}\mbox{.}(2011){Wijesinghe}, {Hopkins}, {Sharp},
  {Gunawardhana}, {Brough}, {Sadler}, {Driver}, {Baldry}, {Bamford}, {Liske},
  {Loveday}, {Norberg}, \& {Peacock}}]{DW2011}
{Wijesinghe} D.~B. {et~al.}, 2011, MNRAS, 410, 2291

\bibitem[{{Wilkins} {et~al}\mbox{.}(2008){Wilkins}, {Trentham}, \&
  {Hopkins}}]{Wilkins2008}
{Wilkins} S.~M., {Trentham} N., {Hopkins} A.~M., 2008, MNRAS, 385, 687

\bibitem[{{Wilman} {et~al}\mbox{.}(2008){Wilman}, {Pierini}, {Tyler}, {McGee},
  {Oemler}, {Morris}, {Balogh}, {Bower}, \& {Mulchaey}}]{Wilman2008}
{Wilman} D.~J. {et~al.}, 2008, ApJ, 680, 1009

\bibitem[{{Wuyts} {et~al}\mbox{.}(2011){Wuyts}, {F{\"o}rster Schreiber},
  {Lutz}, {Nordon}, {Berta}, {Altieri}, {Andreani}, {Aussel}, {Bongiovanni},
  {Cepa}, {Cimatti}, {Daddi}, {Elbaz}, {Genzel}, {Koekemoer}, {Magnelli},
  {Maiolino}, {McGrath}, {P{\'e}rez Garc{\'{\i}}a}, {Poglitsch}, {Popesso},
  {Pozzi}, {Sanchez-Portal}, {Sturm}, {Tacconi}, \& {Valtchanov}}]{Wuyts2011}
{Wuyts} S. {et~al.}, 2011, ApJ, 738, 106

\bibitem[{{Zahid} {et~al}\mbox{.}(2013){Zahid}, {Yates}, {Kewley}, \&
  {Kudritzki}}]{Zahid2013}
{Zahid} H.~J., {Yates} R.~M., {Kewley} L.~J., {Kudritzki} R.~P., 2013, ApJ,
  763, 92

\end{thebibliography}

\bsp

\label{lastpage}

\end{document}